\documentclass[onecolumn,a4paper]{article}
\usepackage{amsfonts}
\usepackage{amsmath}
\usepackage{amssymb}
\usepackage{graphicx}
\usepackage{hyperref}

\newtheorem{theorem}{Theorem}[section]

\newtheorem{remark}[theorem]{Remark}
\numberwithin{equation}{section}
\newtheorem{acknowledgement}{Acknowledgement}

\newtheorem{notation}{Notation}

\providecommand{\keywords}[1]
{
  \small
  \textbf{\textit{Keywords---Quantum mechanics, double-slit experiment, p-adic numbers}} #1
}

\input{tcilatex}

\begin{document}

\title{The $p$-Adic Schr\"{o}dinger Equation and the Two-slit Experiment in
Quantum Mechanics}
\author{W. A. Z\'{u}\~{n}iga-Galindo\thanks{%
The author was partially supported by the Lokenath Debnath Endowed
Professorship.} \\
%EndAName
wilson.zunigagalindo@utrgv.edu \and University of Texas Rio Grande Valley \\
%EndAName
School of Mathematical \& Statistical Sciences\\
One West University Blvd\\
Brownsville, TX 78520, United States.}
\maketitle

\begin{abstract}
$p$-Adic quantum mechanics is constructed from the Dirac-von Neumann axioms
identifying quantum states with square-integrable functions\ on the
$N$-dimensional $p$-adic space, $\mathbb{Q}_{p}^{N}$. This choice is
equivalent to the hypothesis of the discreteness of the space. The time is
assumed to be a real variable. The $p$-Adic quantum mechanics is motivated by
the question: what happens with the standard quantum mechanics if the space
has a discrete nature? The time evolution of a quantum state is controlled by
a nonlocal Schr\"{o}dinger equation obtained from a $p$-adic heat equation by
a temporal Wick rotation. This $p$-adic heat equation describes a particle
performing a random motion in $\mathbb{Q}_{p}^{N}$. The Hamiltonian is a
nonlocal operator; thus, the Schr\"{o}dinger equation describes the evolution
of a quantum state under nonlocal interactions. In this framework, the
Schr\"{o}dinger equation admits complex-valued plane wave solutions, which we
interpret as $p$-adic de Broglie waves. These mathematical waves have all
wavelength $p^{-1}$. In the $p$-adic framework, the double-slit experiment
cannot be explained using the interference of the de Broglie waves. The
wavefunctions can be represented as convergent series in the de Broglie waves,
but the $p$-adic de Broglie waves are just mathematical objects. Only the
square of the modulus of a wave function has a physical meaning as a
time-dependent probability density. These probability densities exhibit
interference patterns similar to the ones produced by `quantum waves.' In the
$p$-adic framework, in the double-slit experiment, each particle goes through
one slit only. The $p$-adic quantum mechanics is an analogue (or model) of the
standard one under the hypothesis of the existence of a Planck length. The
precise connection between these two theories is an open problem. Finally, we
propose the conjecture that the classical de Broglie wave-particle duality is
a manifestation of the discreteness of space-time.

\end{abstract}

\keywords{}
\tableofcontents

%\title[The $p$-Adic Schr\"{o}dinger Equation and the Two-slit Experiment]{The $p$-Adic Schr\"{o}dinger Equation and the Two-slit Experiment in Quantum
%Mechanics}

%\author[Z\'{u}\~{n}iga-Galindo]{W. A. Z\'{u}\~{n}iga-Galindo}
%\address{University of Texas Rio Grande Valley\\
%School of Mathematical \& Statistical Sciences\\
%One West University Blvd\\
%Brownsville, TX 78520, United States}
%\email{wilson.zunigagalindo@utrgv.edu}
%\thanks{The author was partially supported by the Debnath Endowed \ Professorship}
%\subjclass{Primary: 81Q35, 81Q65. Secondary: 26E30}

\section{Introduction}

This paper aims to discuss a model of the double-slit experiment in quantum
mechanics (QM) based on a $p$-adic Schr\"{o}dinger equation. In the Dirac-von
Neumann formulation of QM, the states of a quantum system are vectors of an
abstract complex Hilbert space $\mathcal{H}$, and the observables correspond
to linear self-adjoint operators in $\mathcal{H}$, \cite{Dirac}-\cite{Komech}.
A particular choice of space $\mathcal{H}$ goes beyond the mathematical
formulation and belongs to the domain of the physical practice an intuition,
\cite[Chap. 1, Sect. 1.1]{Berezin et al}. In practice, choosing a particular
Hilbert space also implies the choice of a topology for the space (or
space-time). For instance, if we take $\mathcal{H}=L%
%TCIMACRO{\U{b2}}%
%BeginExpansion
{{}^2}%
%EndExpansion
(\mathbb{R}^{N})$, we are assuming that space ($\mathbb{R}^{N}$) is
continuous, i.e., it is arcwise topological space, which means that there is a
continuous path joining any two points in the space. Let us denote by
$\mathbb{Q}_{p}$ the field of $p$-adic numbers; here, $p$ is a fixed prime
number. The space $\mathbb{Q}_{p}^{N}$ is discrete, i.e., the points and the
empty set are the only connected subsets. The Hilbert spaces $L^{2}%
(\mathbb{R}^{N})$, $L^{2}(\mathbb{Q}_{p}^{N})$ are isometric, since both have
countable orthonormal bases, but the geometries of underlying spaces
($\mathbb{R}^{N}$, $\mathbb{Q}_{p}^{N}$) \ are radically different.

The time evolution of the state of a quantum system is controlled by a
Schr\"{o}dinger equation, which we assumed is obtained from a heat equation by
performing a temporal Wick rotation. This principle is the core of the path
formulation of quantum physics. The $p$-adic heat equations and their
associated Markov processes have been studied extensively in the last
thirty-five years, see, e.g., \cite{V-V-Z}-\cite{Bendikov et al 1}, and the
references therein. Such an equation describes a particle performing a random
motion in a fractal, $\mathbb{Q}_{p}^{N}$. A trajectory is a function
assigning to a time $t\in\left(  0,\infty\right]  $ a point $X(t)\in
\mathbb{Q}_{p}^{N}$. In this framework, any continuous trajectory is a
constant function. Then, in the $p$-adic framework, the motion is just a
sequence of jumps in $\mathbb{Q}_{p}^{N}$.

Here, we use the simplest type, which is%
\begin{equation}
\frac{\partial\Psi(x,t)}{\partial t}+\boldsymbol{D}^{\alpha}\Psi(x,t)=0,\quad
x\in\mathbb{Q}_{p}^{N},\quad t\geq0, \label{Eq_1}%
\end{equation}
where $\boldsymbol{D}^{\alpha}$ is the Taibleson-Vladimirov fractional
derivative, $\alpha>0$; $\boldsymbol{D}^{\alpha}$ is a nonlocal operator. The
discreteness of \ space $\mathbb{Q}_{p}^{N}$ has a strong influence on the
properties of the random motion described by (\ref{Eq_1}). For instance, the
fractional heat equation on $\mathbb{R}^{N}$ is defined by an equation similar
to (\ref{Eq_1}), but with $\alpha\in\left(  0,2\right]  $.

The corresponding free Schr\"{o}dinger equation in natural units is
\begin{equation}
i\frac{\partial\Psi(x,t)}{\partial t}=\boldsymbol{D}^{\alpha}\Psi(x,t)\text{,
}x\in\mathbb{Q}_{p}^{N},\quad t\geq0. \label{Eq_1A}%
\end{equation}
Given $t\geq0$, the Taibleson-Vladimirov fractional derivative of $\Psi(x,t)$
is determined for all the points in $\mathbb{Q}_{p}^{N}$, and not just for
those points close to $x$, then, the equation (\ref{Eq_1A}) describes the
evolution of a quantum state $\Psi(x,t)$ under nonlocal interactions. The
standard QM is compatible with the wave-particle duality. There is a
wavelength (the de Broglie wavelength) manifested in all the objects in
quantum mechanics, which determines the probability density of finding the
object at a given point of the configuration space. The de Broglie wavelength
of a particle is inversely proportional to its momentum. Schr\"{o}dinger
equation (\ref{Eq_1A}) admits complex-valued plane waves, which are wavelets
obtained from a mother wavelet by the action of the scale group of
$\mathbb{Q}_{p}^{N}$. All these `waves' have wavelength $p^{-1}$; thus, they
do not behave like the classical de Broglie waves.

In this framework, we \ develop the mathematical model for the two-slit
experiment described in the abstract. A similar description of the two-slit
experiment was given in \cite{Aharonov et al}: \textquotedblleft Instead of a
quantum wave passing through both slits, we have a localized particle with
nonlocal interactions with the other slit.\textquotedblright\ Here, we obtain
the same conclusion, but in our framework, the nonlocal interactions are a
consequence of the discreteness of the space $\mathbb{Q}_{p}^{N}$.

In the 1930s Bronstein showed that general relativity and quantum mechanics
imply that the uncertainty $\Delta x$ of any length measurement satisfies
\begin{equation}
\Delta x\geq L_{\text{Planck}}:=\sqrt{\frac{\hbar G}{c^{3}}},
\label{Inequality}%
\end{equation}
where $L_{\text{Planck}}$ is the Planck length ($L_{\text{Planck}}%
\approx10^{-33}$ $cm$), \cite{Bronstein}. This inequality establishes an
absolute limitation on length measurements, so the Planck length is the
smallest possible distance that can, in principle, be measured. The choice of
$\mathbb{R}^{N}$ as a model space is not compatible with inequality
(\ref{Inequality}) because, according to the Archimedean axiom valid in
$\mathbb{R}$, any given large segment on a straight line can be surpassed by
successive addition of small segments along the same line. This is a physical
axiom related to the process of measurement, that implies the possibility of
measuring arbitrarily small distances. In the 1980s, Volovich proposed using
$p$-adic numbers instead of real numbers in models at Planck scale, because
the Archimedean axiom is not valid in $\mathbb{Q}_{p}$, \cite{Volovich}.
Intuitively, there are no intervals in $\mathbb{Q}_{p}$ only isolated points
(it$\mathbb{\ }$is a completely disconnected space), so $\mathbb{Q}_{p}$ is
the prototype of a discrete with a very rich mathematical structure. Another
interpretation of Bronstein's inequality drives to quantum gravity,
\cite{Rovelli et al}. Here, it is relevant to mention that $\mathbb{Q}_{p}$ is
invariant under scale transformations of the form $x\rightarrow a+p^{L}x$,
where $a\in\mathbb{Q}_{p}$ and $L\in\mathbb{Z}$. The smallest distance between
two different points in $\mathbb{Q}_{p}$, up to a scale transformation, is
$p^{-1}$. Thus, when we assume $\mathbb{Q}_{p}$ as a model of the space,
$p^{-1}$ is the Planck length.

In the last thirty-five years the $p$-adic quantum mechanics and the $p$-adic
Schr\"{o}\-dinger equations have been studied extensively, see, e.g.,
\cite{Beltrameti et al}-\cite{Aniello et al}, among many available references.
There are at least three different types of $p$-adic Schr\"{o}\-dinger
equations. In the first type, the wave functions are complex-valued and the
space-time is $\mathbb{Q}_{p}^{N}\times\mathbb{R}$; in the second type, the
wave functions are complex-valued and the space-time is $\mathbb{Q}_{p}%
^{N}\times\mathbb{Q}_{p}$; in the third type, the wave functions are $p$-adic-
valued and the space-time is $\mathbb{Q}_{p}^{N}\times\mathbb{Q}_{p}$.
Time-independent Schr\"{o}\-dinger equations have been intensively studied, in
particular, the spectra of the corresponding Schr\"{o}\-dinger operators, see,
e.g., \cite[Chap. 2, Sect. X, \ and \ Chap. 3, Sects. \ XI, XII]{V-V-Z},
\cite[Chap. 3]{Kochubei}. Feynman-Kac formulas for a large class of potentials
involving some generalizations of the one-dimensional Taibleson-Vladimirov
operator were established in \cite{Ismagilov}, \cite{Blair},
\cite{Varadarjan1}.

In the last thirty-five years the connection between $q$-oscillator algebras
and quantum physics have been studied intensively, see, e.g.,
\cite{arefeve-volovich}-\cite{Zhang-2}, and the references therein. From the
seminal work of Biedenharn \cite{Biedenharn} and Macfarlane \cite{Macfarlane},
it was clear that the $q$-analysis, \cite{Ernst}, \cite{Kac}-\cite{Klimyk},
plays a central role in the representation of $q$-oscillator algebras which in
turn have a deep physical meaning. In particular, the $q$-deformation of the
Heisenberg algebra drives naturally to several types of $q$-deformed
Schr\"{o}dinger equations, see e.g. \cite{arefeve-volovich}, \cite{Lavagno}%
-\cite{Lavagno2}, \cite{Wess-Zumino}-\cite{Zhang-2}. In
\cite{arefeve-volovich} Aref'eva and Volovich pointed out the existence of
deep analogies between $p$-adic and $q$-analysis, and between $q$-deformed
quantum mechanics and $p$-adic quantum mechanics. In \cite{Zuniga-QM}, the
author \ shows that many quantum models constructed using $q$-oscillator
algebras are non-Archimedean models, in particular, $p$-adic quantum models.
Together \cite{arefeve-volovich} and \cite{Zuniga-QM} suggest the existence of
a deep mathematical connection between $p$-adic QM and quantum groups.

In the 1930s, Weyl and later in 1950s Schwinger, considered the approximation
of quantum systems using unconventional rings and fields, \cite{Schwinger-1}%
-\cite{VourdasBook}. The finite approximation of quantum systems has become a
very relevant research area due to its applications to quantum optics, quantum
computing, two-dimensional electron systems in magnetic fields and the
magnetic translation group, the quantum Hall effect, etc., see
\cite{VourdasBook}, and the references therein. A correspondence between
Euclidean quantum fields and neural networks has recently been proposed. Such
correspondence is expected to explain how deep learning architectures work.
The author and some of his collaborators have developed a $p$-adic counterpart
of this correspondence. Finite approximations of $p$-adic Euclidean quantum
field theories correspond to Boltzmann machines with a hierarchical
architecture, \cite{PTEP-2023}-\cite{PHysicaA-2023}.

\section{The Dirac-von Neumann formulation of QM}

In the Dirac-Von Neumann formulation of QM, to every isolated quantum system
there is associated a separable complex Hilbert space $\mathcal{H}$ called the
space of states. The Hilbert space of a composite system is the Hilbert space
tensor product of the state spaces associated with the component systems. The
states of a quantum system are described by non-zero vectors from
$\mathcal{H}$. Two vectors describe the same state if and only if they differ
by a non-zero complex factor. Each observable corresponds to a unique linear
self-adjoint operator in $\mathcal{H}$. \ The most important observable of a
quantum system is its energy. We denote the corresponding operator by
$\boldsymbol{H}$. Let $\Psi_{0}\in\mathcal{H}$ be the state at time $t=0$ of a
certain quantum system. Then at any time $t$ the system is represented by the
vector $\Psi\left(  t\right)  =\boldsymbol{U}_{t}\Psi_{0}$, where
$\boldsymbol{U}_{t}$\ is a unitary operator called the evolution operator. The
evolution operators $\left\{  \boldsymbol{U}_{t}\right\}  _{t\geq0}$ form a
strongly continuous one-parameter unitary group on $\mathcal{H}$. The vector
function $\Psi\left(  t\right)  $ is differentiable if $\Psi\left(  t\right)
$ is contained in the domain $D(\boldsymbol{H})$ of $\boldsymbol{H}$, which
happens \ if at $t=0$, $\Psi_{0}\in D(\boldsymbol{H})$, and in this case the
time evolution of $\Psi\left(  t\right)  $ is controlled by the
Schr\"{o}dinger equation
\[
i\frac{\partial}{\partial t}\Psi\left(  t\right)  =\boldsymbol{H}\Psi\left(
t\right)  \text{, }%
\]
where $i=\sqrt{-1}$ and the Planck constant is assumed to be one. The
evolution operators $\left\{  \boldsymbol{U}_{t}\right\}  _{t\geq0}$ form a
strongly continuous group generated by $-i\boldsymbol{H}$ , that is%
\[
\Psi\left(  t\right)  =\boldsymbol{U}_{t}\Psi_{0}=e^{-it\boldsymbol{H}}%
\Psi_{0},\text{ }t\geq0\text{.}%
\]
For an in-depth discussion the reader may consult \cite{Dirac}-\cite{Komech}.

In standard QM, the states of quantum systems are functions from spaces of
type $L^{2}(\mathbb{R}^{N})$ or $\mathbb{C}^{N}$. In the first case, the wave
functions take the form%
\[
\Psi\left(  x,t\right)  :\mathbb{R}^{N}\times\mathbb{R}_{+}\rightarrow
\mathbb{C}\text{,}%
\]
where $x\in\mathbb{R}^{N}$, and $t\in\mathbb{R}_{+}:=\left\{  t\in
\mathbb{R};t\geq0\right\}  $. This choice implies that the space is
continuous, i.e., given two different points $x_{0}$, $x_{1}\in\mathbb{R}^{N}$
there exists a continuous curve $X\left(  t\right)  :\left[  a,b\right]
\rightarrow\mathbb{R}^{N}$ such that $X\left(  a\right)  =x_{0}$, $X\left(
b\right)  =x_{1}$. The Dirac-von Neumann formulation of QM does not rule out
the possibility of choosing a discrete space. By a discrete space, we mean a
completely disconnected topological space, which is a topological space where
the connected components are the points and the empty set. In a such space
does not exist a continuous curve joining two different points.

\section{Wick rotation and the heat equation}
Let us denote by $\boldsymbol{H}_{0}$ the standard $N$-dimensional Laplacian.
The Wick rotation $t\rightarrow it$ changes the heat equation%
\begin{equation}
\frac{\partial}{\partial t}\Psi\left(  x,t\right)  =-\boldsymbol{H}_{0}%
\Psi\left(  x,t\right)  \label{Heat_Equation}%
\end{equation}
into the Schr\"{o}dinger equation%
\[
i\frac{\partial}{\partial t}\Psi\left(  x,t\right)  =\boldsymbol{H}_{0}%
\Psi\left(  x,t\right)  .
\]
The fundamental solution of the heat equation is the transition density of a
Markov process, i.e., $e^{-t\boldsymbol{H}_{0}}$ is a Feller semigroup, see,
e.g., \cite[Part I, Section 3]{Taira}. This observation is at the core of the
path formulation of QM, see, e.g., \cite{Zee}.

\section{The field of $p$-adic numbers}

From now on, we use $p$ to denote a fixed prime number and the letter
$\mathsf{p}$\ to denote the momentum of a particle. Only these two letters
will appear together in Section \ref{Section_7}.

\begin{figure}[ptb]
\begin{center}
\includegraphics[height=2.7942in,width=4.9485in,angle=0]{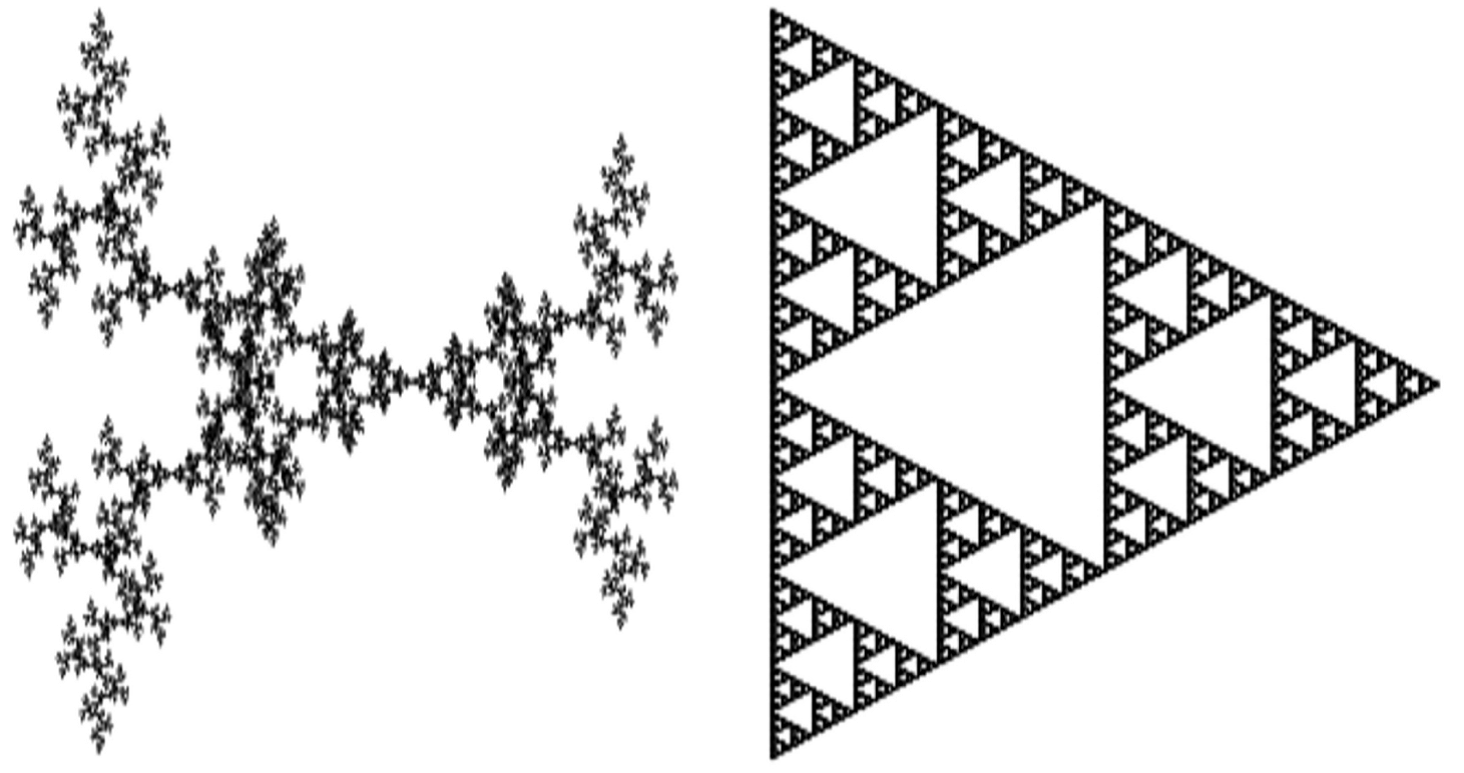}
\end{center}
\caption{Based upon \protect\cite{Chistyakov}, construct an embedding $%
\mathfrak{f}:\mathbb{Z}_{\wp }\rightarrow\mathbb{R}^{2}$. The figure shows
the images of $\mathfrak{f}(\mathbb{Z}_{2})$ and $\mathfrak{f}(\mathbb{Z}%
_{3})$. This computation requires a truncation of the $\wp$-adic integers.
We use $\mathbb{Z}_{2}/2^{14}\mathbb{Z}_{2}$ and $\mathbb{Z}_{3}/3^{10}%
\mathbb{Z}_{3}$, respectively. Taken from \protect\cite{PTEP-2023}.}
\label{Figure 1}
\end{figure}

Any non-zero $p$-adic number $x$ has a unique expansion of the form%
\begin{equation}
x=x_{-k}p^{-k}+x_{-k+1}p^{-k+1}+\ldots+x_{0}+x_{1}p+\ldots,\text{ }
\label{p-adic-number}%
\end{equation}
with $x_{-k}\neq0$, where $k$ is an integer, and the $x_{j}$s \ are numbers
from the set $\left\{  0,1,\ldots,p-1\right\}  $. The set of all possible
sequences form the (\ref{p-adic-number}) constitutes the field of $p$-adic
numbers $\mathbb{Q}_{p}$. There are natural field operations, sum and
multiplication, on series of form (\ref{p-adic-number}). There is also a norm
in $\mathbb{Q}_{p}$ defined as $\left\vert x\right\vert _{p}=p^{k}$, for a
nonzero $p$-adic number $x$. The field of $p$-adic numbers with the distance
induced by $\left\vert \cdot\right\vert _{p}$ is a complete ultrametric space.
The ultrametric property refers to the fact that $\left\vert x-y\right\vert
_{p}\leq\max\left\{  \left\vert x-z\right\vert _{p},\left\vert z-y\right\vert
_{p}\right\}  $ for any $x$, $y$, $z$ in $\mathbb{Q}_{p}$. The $p$-adic
integers which are sequences of the form (\ref{p-adic-number}) with $-k\geq0$.
All these sequence constitute the unit ball $\mathbb{Z}_{p}$. As a topological
space $\mathbb{Q}_{p}$\ is homeomorphic to a Cantor-like subset of the real
line, see, e.g., \cite{V-V-Z}, \cite{Chistyakov}, \cite{Alberio et al}. See
Figure \ref{Figure 1}.

\begin{figure}[ptb]
\begin{center}
\includegraphics[height=2.1024in,width=3.717in,angle=0]{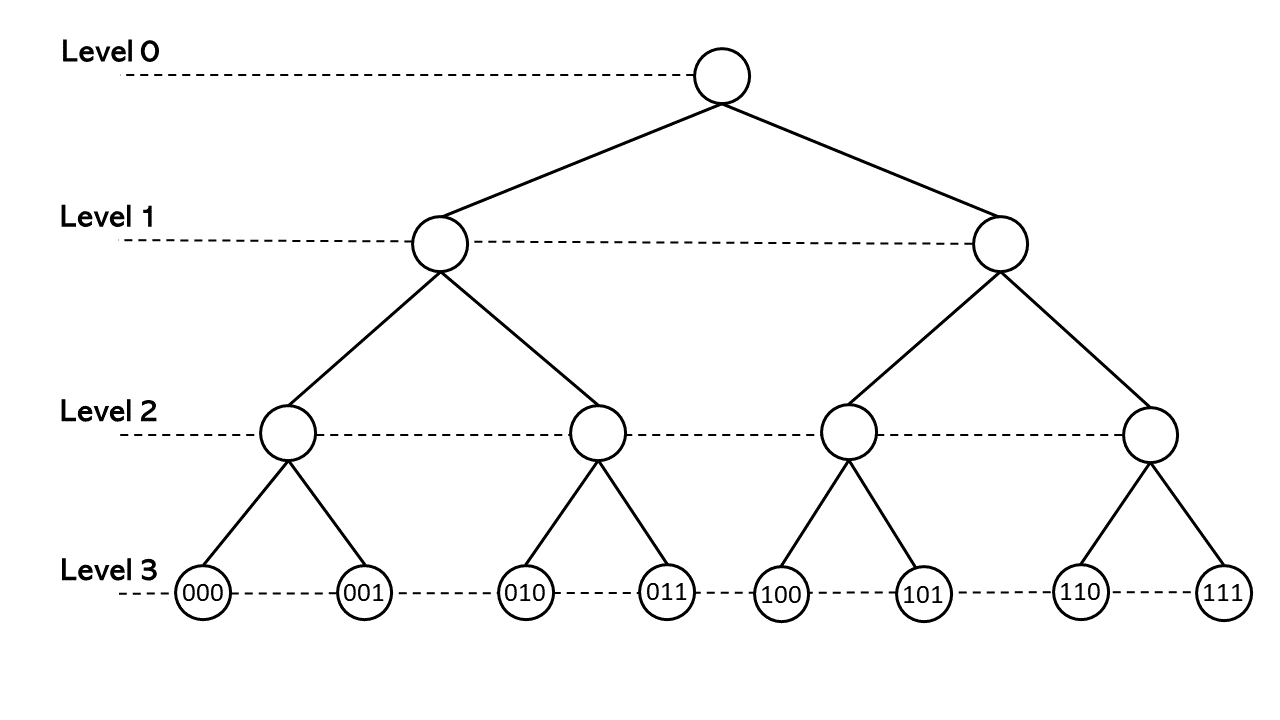}
\end{center}
\caption{The rooted tree associated with the group $\mathbb{Z}_{2}/2^{3}%
\mathbb{Z}_{2}$. The elements of $\mathbb{Z}_{2}/2^{3}\mathbb{Z}_{2}$ have
the form $i=i_{0}+i_{1}2+i_{2}2^{2}$,$\;i_{0}$, $i_{1}$, $i_{2}\in\{0,1\}$.
The distance satisfies $-\log_{2}\left\vert i-j\right\vert _{2}=$level of
\noindent the first common ancestor of $i$, $j$. Taken from \protect\cite%
{PTEP-2023}.}
\label{Figure 2}
\end{figure}

There is a natural truncation operation on $p$-adic integers:
\[
x=%
%TCIMACRO{\dsum \limits_{k=0}^{\infty}}%
%BeginExpansion
{\displaystyle\sum\limits_{k=0}^{\infty}}
%EndExpansion
x_{k}p^{k}\rightarrow x_{0}+x_{1}p+\ldots+x_{l-1}p^{l-1}\text{, \ }l\geq1.
\]
The set all truncated integers mod $p^{l}$ is denoted as $G_{l}=\mathbb{Z}%
_{p}/p^{l}\mathbb{Z}_{p}$. This set is \ a rooted tree with $l$ levels, see
Figure \ref{Figure 2}. The unit ball $\mathbb{Z}_{p}$ (which is the inverse
limit of the $G_{l}$s) is an infinite rooted tree with fractal structure, see
Figure \ref{Figure 1}.%

The unit ball $\mathbb{Z}_{p}$ (which is the inverse limit of the $G_{l}$s)
is an infinite rooted tree with fractal structure, see Figure \ref{Figure 1}.

We extend the $p$-adic norm to $\mathbb{Q}_{p}^{N}$ by taking%
\[
||x||_{p}:=\max_{1\leq i\leq N}|x_{i}|_{p},\qquad\text{for }x=(x_{1}%
,\dots,x_{N})\in\mathbb{Q}_{p}^{N}.
\]
We define $ord(x)=\min_{1\leq i\leq N}\{ord(x_{i})\}$, then $||x||_{p}%
=p^{-ord(x)}$.\ The metric space $\left(  \mathbb{Q}_{p}^{N},||\cdot
||_{p}\right)  $ is a complete ultrametric space.

A function $\varphi:\mathbb{Q}_{p}^{N}\rightarrow\mathbb{C}$ is called locally
constant, if for any $a\in\mathbb{Q}_{p}^{N}$, there is an integer $l=l(a)$,
such that
\[
\varphi\left(  a+x\right)  =\varphi\left(  a\right)  \text{ for any }%
||x||_{p}\leq p^{l}.
\]
The set of functions for which $l=l\left(  \varphi\right)  $ depends only on
$\varphi$ form a $\mathbb{C}$-vector space denoted as $\mathcal{U}%
_{loc}\left(  \mathbb{Q}_{p}^{N}\right)  $. We call $l\left(  \varphi\right)
$ the exponent of local constancy. If $\varphi\in\mathcal{U}_{loc}\left(
\mathbb{Q}_{p}^{N}\right)  $ has compact support, we say that $\varphi$ is a
test function. We denote by $\mathcal{D}(\mathbb{Q}_{p}^{N})$ the complex
vector space of test functions. There is a natural integration theory so that
$\int_{\mathbb{Q}_{p}^{N}}\varphi\left(  x\right)  dx$ gives a well-defined
complex number. The measure $d^{N}x$ is the Haar measure of $\mathbb{Q}%
_{p}^{N}$. In the Appendix (Section \ref{Appendix}), we give a quick review of
the basic aspects of the $p$-adic analysis required here.

The QM in the sense of the Dirac-von Neumann can be formulated on the Hilbert
space%
\[
L^{2}(\mathbb{Q}_{p}^{N}):=L^{2}(\mathbb{Q}_{p}^{N},d^{N}x)=\left\{
f:\mathbb{Q}_{p}^{N}\rightarrow\mathbb{C};\left\Vert f\right\Vert _{2}=\left(
%
%TCIMACRO{\dint \limits_{\mathbb{Q}_{p}^{N}}}%
%BeginExpansion
{\displaystyle\int\limits_{\mathbb{Q}_{p}^{N}}}
%EndExpansion
\left\vert f\left(  x\right)  \right\vert ^{2}d^{N}x\right)  ^{\frac{1}{2}%
}<\infty\right\}  .
\]
This space has a countable orthonormal basis, and thus $L^{2}(\mathbb{Q}%
_{p}^{N},d^{N}x)$ is isometric to $l_{2}(\mathbb{C})$. On the other hand,
$L^{2}(\mathbb{R}^{N},d^{N}x)$, here $d^{N}x$ denotes the Lebesgue measure of
$\mathbb{R}^{N}$, is also isometric to $l_{2}(\mathbb{C})$, and thus
$L^{2}(\mathbb{R}^{N},d^{N}x)$ is isometric to $L^{2}(\mathbb{Q}_{p}^{N}%
,d^{N}x)$. However, the first choice $L^{2}(\mathbb{R}^{N},d^{N}x)$ implies
that the space ($\mathbb{R}^{N}$) is continuous, while in the second choice
$L^{2}(\mathbb{Q}_{p}^{N},d^{N}x)$, the space ($\mathbb{Q}_{p}^{N}$) is a
completely disconnected.

\section{$p$-Adic \ numbers in physics}

Using $p$-adic numbers in physics raises several new questions. Here, we
discuss two which are relevant in this work.

\subsection{What is the role of $p$?}

For the sake of simplicity, in order to discuss this matter we take $N=1$. The
transformations of type%
\[
\rho_{a,L}:x\rightarrow a+p^{L}x\text{, with }a\in\mathbb{Q}_{p}\text{, }%
L\in\mathbb{Z},
\]
constitute a scale group for the set $\mathbb{Q}_{p}\setminus\left\{
0\right\}  $. This set is self-similar, i.e.,
\[
\mathbb{Q}_{p}\setminus\left\{  0\right\}  =%
%TCIMACRO{\dbigcup \limits_{a,L}}%
%BeginExpansion
{\displaystyle\bigcup\limits_{a,L}}
%EndExpansion
\rho_{a,L}\left(  \mathbb{Q}_{p}\setminus\left\{  0\right\}  \right)  .
\]
Given two different points $x_{1}$, $x_{2}$ in $\mathbb{Q}_{p}\setminus
\left\{  0\right\}  $, we say that they are equivalent, denoted as $x_{1}\sim$
$x_{2}$, if $\rho_{a,L}\left(  x_{1}\right)  =x_{2}$, for some $\rho_{a,L}$.
Then, any non-zero $p$-adic number%
\[
x=p^{ord(x)}%
%TCIMACRO{\dsum \limits_{k=0}^{\infty}}%
%BeginExpansion
{\displaystyle\sum\limits_{k=0}^{\infty}}
%EndExpansion
x_{k}p^{k}=:p^{ord(x)}ac\left(  x\right)  \text{, with }x_{0}\neq0\text{ and
}ac\left(  x\right)  \in\mathbb{Z}_{p}^{\times}\text{,}%
\]
is equivalent to a point on the unit sphere $\mathbb{Z}_{p}^{\times}=\left\{
x\in\mathbb{Z}_{p};\left\vert x\right\vert _{p}=1\right\}  $. Now,
\[
\mathbb{Z}_{p}^{\times}=%
%TCIMACRO{\dbigsqcup \limits_{j_{0}=1}^{p-1}}%
%BeginExpansion
{\displaystyle\bigsqcup\limits_{j_{0}=1}^{p-1}}
%EndExpansion%
%TCIMACRO{\dbigsqcup \limits_{j_{1}=0}^{p-1}}%
%BeginExpansion
{\displaystyle\bigsqcup\limits_{j_{1}=0}^{p-1}}
%EndExpansion
\left(  j_{0}+j_{1}p+p^{2}\mathbb{Z}_{p}\right)  \text{,}%
\]
where $%
%TCIMACRO{\tbigsqcup }%
%BeginExpansion
{\textstyle\bigsqcup}
%EndExpansion
$ denotes\ disjoint union. By using a suitable transformation $\rho_{a,L}$ any
element of the set $j_{0}+j_{1}p+p^{2}\mathbb{Z}_{p}$ is equivalent to
$j_{0}+j_{1}p$. These elements form a rooted tree $T_{j_{0}}$, with root
$j_{0}$, and one layer. The vertices on this layer correspond with the points
of the form $j_{0}+j_{1}p$. In conclusion, $\mathbb{Q}_{p}\setminus\left\{
0\right\}  $ is a self-similar set with a tree-like structure, having
the\ forest $%
%TCIMACRO{\tbigsqcup \nolimits_{j_{0}=1}^{p-1}}%
%BeginExpansion
{\textstyle\bigsqcup\nolimits_{j_{0}=1}^{p-1}}
%EndExpansion
T_{j_{0}}$ as a fundamental domain for the action of the scale group. Then, up
to a scale transformation, \ the smallest distance \ between two points in
$\mathbb{Q}_{p}\setminus\left\{  0\right\}  $ is
\[
p^{-1}=\left\vert j_{0}-\left(  j_{0}+j_{1}p\right)  \right\vert _{p}.
\]
We warn the reader about the following facts. If we consider the number $p$ as
a real number $p=\left\vert p\right\vert $. But, if we consider $p$ as a
$p$-adic number, $p\neq\left\vert p\right\vert _{p}=p^{-1}$. This phenomenon
causes big trouble when using $p$-adic numbers in physics. We introduce
\textit{the principle of the inaccessibility of }$p$\textit{-adic numbers}.
Only the $p$-adic norm of the result of a computation using $p$-adic numbers
can be compared against some measurable physical quantity. The physical
interpretation of the real \ values of $p$-adic \ numbers is an open problem.

\subsection{\label{Section_Motion}Motion in a $p$-adic space}

The product topology of $\mathbb{Q}_{p}^{N}\times\mathbb{R}_{+}$ imposes
serious restrictions to the notion of motion. In classical physics the
trajectory of a particle in $\mathbb{R}^{N}$ is a continuous curve contained
in $\mathbb{R}^{N}$, i.e.,
\[%
\begin{array}
[c]{ccc}%
\left[  0,\infty\right)  & \rightarrow & \mathbb{R}^{N}\\
&  & \\
t & \rightarrow & X(t).
\end{array}
\]
In the $p$-adic case, the trajectories have the form%
\[%
\begin{array}
[c]{ccc}%
\left[  0,\infty\right)  & \rightarrow & \mathbb{Q}_{p}^{N}\\
&  & \\
t & \rightarrow & Y(t).
\end{array}
\]
If the $Y$ is continuous, the image $Y(\left[  0,\infty\right)  )$ should be a
connected subset of $\mathbb{Q}_{p}^{N}$, since the only connected subsets of
$\mathbb{Q}_{p}^{N}$ are the empty set, and the points, the function $Y$
should be constant. Therefore, the non-trivial trajectories are discontinuous
curves. This mathematical result can be reformulated as the motion of a
particle in $\mathbb{Q}_{p}^{N}$ is a sequence of jumps.

Suppose that $Y(t)\in\mathbb{Q}_{p}^{N}$ , $t\in\left[  0,\infty\right)  $, is
the trajectory of a particle in $\mathbb{Q}_{p}^{N}$. Consider a function $F:$
$\mathbb{Q}_{p}^{N}\rightarrow\mathbb{C}$, then $F\left(  Y(t)\right)
\in\mathbb{C}$ \textit{is not the trajectory of particle in} $\mathbb{Q}%
_{p}^{N}$. The function $F\left(  Y(t)\right)  $ is just a picture that may
capture relevant features of a motion in $\mathbb{Q}_{p}^{N}$. For instance,
linear combinations of functions of type $F\left(  Y(t)\right)  =\exp\left(
2\pi if\left\Vert Y(t)\right\Vert \right)  $, $f\in\mathbb{R}$, may produced
\ interference patterns in $\mathbb{C=R}^{2}$, but we cannot interpret theses
patterns as produced by\ the interference of waves in $\mathbb{Q}_{p}^{N}$.

\section{$p$-Adic heat equations and Markov processes}

In this section, we review quickly the\ basic aspects of the $p$-adic heat
equations. For an in-depth \ discussion the reader may consult
\cite{Zuniga-LNM-2016}-\cite{Bendikov et al 1}, and the references therein.

We denote by $\mathcal{C}(\mathbb{Q}_{p}^{N})$ the $\mathbb{C}$-vector space
of continuous functions defined on $\mathbb{Q}_{p}^{N}$. The
Taibleson-Vladimirov pseudodifferential operator $\boldsymbol{D}^{\alpha}$,
$\alpha>0$, is defined as%
\begin{equation}%
\begin{array}
[c]{cccc}%
\boldsymbol{D}^{\alpha}: & \mathcal{D}(\mathbb{Q}_{p}^{N}) & \rightarrow &
L^{2}(\mathbb{Q}_{p}^{N})\cap\mathcal{C}(\mathbb{Q}_{p}^{N})\\
&  &  & \\
& \varphi(x) & \rightarrow & \boldsymbol{D}^{\alpha}\varphi(x)=\mathcal{F}%
_{\xi\rightarrow x}^{-1}\left(  ||\xi||_{p}^{\alpha}\mathcal{F}_{x\rightarrow
\xi}\left(  \varphi\right)  \right)  ,
\end{array}
\label{Taibleson_Vladimirov_Operator}%
\end{equation}
where $\mathcal{F}$ denotes the Fourier transform, see the Appendix for the
definition and further details about the Fourier transform on $\mathbb{Q}%
_{p}^{N}$. This operator admits an extension of the form
\[
\left(  \boldsymbol{D}^{\alpha}\varphi\right)  \left(  x\right)
=\frac{1-p^{\alpha}}{1-p^{-\alpha-N}}\int\limits_{\mathbb{Q}_{p}^{N}}%
||y||_{p}^{-\alpha-N}(\varphi(x-y)-\varphi(x))\,d^{N}y
\]
to the space of locally constant functions $\varphi(x)$ satisfying
\[
\int\limits_{||x||_{p}\geq1}||x||_{p}^{-\alpha-N}|\varphi(x)|\,d^{N}x<\infty.
\]
In particular, the Taibleson-Vladimirov derivative of any order of a constant
function is zero.

The equation%
\[
\frac{\partial u(x,t)}{\partial t}+\boldsymbol{D}^{\alpha}u(x,t)=0,\quad
x\in\mathbb{Q}_{p}^{N},\quad t\geq0,
\]
is called the $p$-adic heat equation. The\ heat kernel attached to the symbol
$a||\xi||_{p}^{\alpha}$, $a>0$,$\,$ is defined as
\begin{equation}
Z(x,t)=\int\limits_{\mathbb{Q}_{p}^{N}}\chi_{p}(-x\cdot\xi)e^{-t||\xi
||_{p}^{\alpha}}\,d^{N}\xi\text{, for }t>0\text{ \ and }x\in\mathbb{Q}_{p}%
^{N}. \label{Heat_Kernel}%
\end{equation}
Since $e^{-at||\cdot||_{p}^{\alpha}}\in L^{1}\left(  \mathbb{Q}_{p}%
^{N}\right)  $ for $t>0$, $Z(x,t)$ is a continuous function in $x\in
\mathbb{Q}_{p}^{N}$ for $t>0$. The heat kernel has the following properties:

\noindent(i) $Z(x,t)\geq0$, for all $x\in\mathbb{Q}_{p}^{N}$,$\quad
t\in(0,\infty)$;

\noindent(ii) $%
%TCIMACRO{\tint \nolimits_{\mathbb{Q}_{p}^{N}}}%
%BeginExpansion
{\textstyle\int\nolimits_{\mathbb{Q}_{p}^{N}}}
%EndExpansion
Z\left(  x,t\right)  $ $d^{N}x=1$, for any $t>0$;

\noindent(iii) if $\varphi\in\mathcal{D}\left(  \mathbb{Q}_{p}^{N}\right)  $,
then $\lim_{\left(  x,t\right)  \rightarrow\left(  x_{0},0\right)  }%
%TCIMACRO{\tint \nolimits_{\mathbb{Q}_{p}^{N}}}%
%BeginExpansion
{\textstyle\int\nolimits_{\mathbb{Q}_{p}^{N}}}
%EndExpansion
Z\left(  x-\eta,t\right)  \varphi\left(  \eta\right)  d^{N}\eta=\varphi\left(
x_{0}\right)  $;

\noindent(iv) $Z\left(  x,t+t^{\prime}\right)  =%
%TCIMACRO{\tint \nolimits_{\mathbb{Q}_{p}^{N}}}%
%BeginExpansion
{\textstyle\int\nolimits_{\mathbb{Q}_{p}^{N}}}
%EndExpansion
Z\left(  x-y,t\right)  Z\left(  y,t^{\prime}\right)  d^{N}y$, for $t$,
$t^{\prime}>0$,

\noindent see \cite[Chaper 2, \ Theorem 13]{Zuniga-LNM-2016}. In addition, the
heat kernel $Z\left(  x,t\right)  $ is a transition density of a time
homogeneous Markov process with discontinuous paths, \cite[Chaper 2, \ Theorem
16]{Zuniga-LNM-2016}.

In $\mathbb{R}^{N}$, the fractional Laplacian is defined as\ $\widehat{\left(
-\Delta f\right)  ^{\beta}}=\left\vert \xi\right\vert _{\mathbb{R}}^{\beta
}\widehat{f}$, for $\beta\in\left[  0,2\right]  $, on the \ Schwartz space,
where $\widehat{\cdot}$ denotes the Fourier transform on $\mathbb{R}^{N}$, see
\cite[Chaper 4]{Zuniga-LNM-2016}, and the references therein. The fractional
heat equation in $\mathbb{R}^{N}$ is%
\[
\frac{\partial u(x,t)}{\partial t}+a\left(  -\Delta f\right)  ^{\beta
}u(x,t)=0,\quad x\in\mathbb{R}^{N},\quad t\geq0,
\]
with $a>0$. There are `more' $p$-adic heat equations than Archimedean ones,
see \cite[Chapter 4]{Zuniga-LNM-2016}. For instance, the $p$-adic heat
equations \ with variable coefficients presented in \cite[Chapter
3]{Zuniga-LNM-2016} do not have Archimedean counterparts. All the $p$-adic
Laplacians are non-local operators, in addition, they can be combined with the
Archimedean fractional Laplacian into an adelic operator, and the
corresponding adelic heat equation is associated with a Markov process whose
state space the ring of adeles, see \cite[Chapter 4, Theorem 123]%
{Zuniga-LNM-2016}.

\section{$p$-Adic Schr\"{o}dinger equations}

A large class $p$-adic quantum mechanics is obtained from the Dirac-von
Neumann formulation by taking $\mathcal{H}=L^{2}(\mathbb{Q}_{p}^{N})$. A
specific Schr\"{o}dinger equation controls the dynamic evolution of the
quantum states in each particular QM. Such an equation is obtained from a
$p$-adic heat by the Wick rotation. We select the simplest possible heat
equation:%
\begin{equation}
\frac{\partial\Psi(x,t)}{\partial t}=-\boldsymbol{D}^{\alpha}\Psi(x,t),\quad
x\in\mathbb{Q}_{p}^{N},\quad t\geq0, \label{Heat_Equation_2}%
\end{equation}
where $\boldsymbol{D}^{\alpha}$ is the Taibleson-Vladimirov
pseudo-differential operator. By performing the Wick rotation $t\rightarrow
it$, where $i=\sqrt{-1}$, (\ref{Heat_Equation_2}) becomes%
\begin{equation}
i\frac{\partial\Psi(x,t)}{\partial t}=\boldsymbol{D}^{\alpha}\Psi(x,t),\quad
x\in\mathbb{Q}_{p}^{N},\quad t\geq0. \label{Schrodinger_Equation}%
\end{equation}
We propose the following Schr\"{o}dinger equation for a single nonrelativistic
particle with mass $m$ moving in $\mathbb{Q}_{p}^{N}$ under a potential
$V(x,t)$:%
\begin{equation}
i\hbar\frac{\partial\Psi(x,t)}{\partial t}=\left\{  \frac{\hbar^{2}}%
{2m}\boldsymbol{D}^{\alpha}+V(x,t)\right\}  \Psi(x,t)\text{, }x\in
\mathbb{Q}_{p}^{N},\quad t\geq0. \label{Schrodinger_Equation_1}%
\end{equation}
If $\frac{\hbar^{2}}{2m}=1$ and $V(x,t)=0$, then equation
(\ref{Schrodinger_Equation_1}) becomes (\ref{Schrodinger_Equation}). For the
rest of the section, we assume that $\frac{\hbar^{2}}{2m}=1$ and that $V(x)$
is time-independent.\ The Schr\"{o}dinger operators $\boldsymbol{D}^{\alpha
}+V(x)$ has been extensively studied, see, e.g., \cite[Chapter 3]{Kochubei},
\cite[Chapter 2.]{V-V-Z}, \cite{Bendikov et al}.

\subsection{Time evolution operators}

We recall that a function $V:\mathbb{Q}_{p}^{N}\rightarrow\mathbb{R}$ is
called locally bounded, if for any $x_{0}\in\mathbb{Q}_{p}^{N}$, there exists
a neighborhood $\mathcal{N}$ of $x_{0}$ and a positive constant $M$ such that
\[
\left\vert f\left(  x\right)  \right\vert \leq M\text{ for all }%
x\in\mathcal{N}.
\]
Let $V:\mathbb{Q}_{p}^{N}\rightarrow\mathbb{R}$ be a locally bounded,
measurable function. Define the operator $\boldsymbol{H}^{\prime}$ on
$L^{2}(\mathbb{Q}_{p}^{N})$ with domain $D(\boldsymbol{H}^{\prime
})=\mathcal{D}(\mathbb{Q}_{p}^{N})$ setting
\[
\boldsymbol{H}^{\prime}\varphi\left(  x\right)  =\left(  \boldsymbol{D}%
^{\alpha}+V(x)\right)  \varphi\left(  x\right)  .
\]
Since $\mathcal{D}(\mathbb{Q}_{p}^{N})$ is dense in $L^{2}(\mathbb{Q}_{p}%
^{N})$ and $\boldsymbol{D}^{\alpha}\varphi\left(  x\right)  $, $V(x)\varphi
\left(  x\right)  \in L^{2}(\mathbb{Q}_{p}^{N})$, see
(\ref{Taibleson_Vladimirov_Operator}), $\boldsymbol{H}^{\prime}$ is a densely
defined and symmetric operator. We denote by $\boldsymbol{H}$ its closure.
This operator \ is self-adjoint \cite[Theorem 3.2]{Kochubei}. Then, by Stone's
theorem, $e^{-it\boldsymbol{H}}$ is a unitary group on $L^{2}(\mathbb{Q}%
_{p}^{N})$ \ and
\[
\Psi(x,t)=e^{-it\boldsymbol{H}}\Psi_{0}(x)\text{, for }t\in\mathbb{R}\text{,
and }\Psi_{0}(x)\in L^{2}(\mathbb{Q}_{p}^{N})\text{,}%
\]
is the solution of the Cauchy problems attached to the $p$-adic
Schr\"{o}dinger equation with initial datum $\Psi_{0}$. In \cite{Ismagilov},
\cite{Blair}, \cite{Varadarjan1}, Feynman-Kac formulas for the semigroups
$e^{-it\boldsymbol{H}}$ for large class of potentials were established.

\subsection{A Cauchy problem}

We now consider the following Cauchy problem:%
\begin{equation}
\left\{
\begin{array}
[c]{ll}%
\Psi(\cdot,t)\in\mathcal{C}\left(  \left[  0,\infty\right)  ,\mathcal{D}%
(\mathbb{Q}_{p}^{N})\right)  ; & \Psi(\cdot,t)\in\mathcal{C}^{1}\left(
\left[  0,\infty\right)  ,L^{2}(\mathbb{Q}_{p}^{N})\right) \\
& \\
\frac{\partial\Psi(x,t)}{\partial t}=-i\boldsymbol{D}^{\alpha}\Psi(x,t), &
x\in\mathbb{Q}_{p}^{N}\text{, }t\geq0\\
& \\
\Psi(x,0)=\Psi_{0}(x)\in\mathcal{D}(\mathbb{Q}_{p}^{N}). &
\end{array}
\right.  \label{Cauchy_Problem_1}%
\end{equation}
By taking Fourier transform in the space variables, the solution of the
evolution equation in (\ref{Cauchy_Problem_1}) is given by%
\[
\widehat{\Psi}(\xi,t)=\widehat{\Psi}_{0}(\xi)e^{-it||\xi||_{p}^{\alpha}},
\]
now, since $\widehat{\Psi}_{0}(\xi)\in\mathcal{D}(\mathbb{Q}_{p}^{N})$ and
$e^{-it||\xi||_{p}^{\alpha}}\in\mathcal{C}(\mathbb{Q}_{p}^{N})$, it verifies
that
\[
\widehat{\Psi}_{0}(\xi)e^{-it||\xi||_{p}^{\alpha}}\in L^{1}(\mathbb{Q}_{p}%
^{N})\cap L^{2}(\mathbb{Q}_{p}^{N})\text{, for }t\geq0,
\]
and consequently%
\begin{equation}
\Psi(x,t)=%
%TCIMACRO{\dint \limits_{\mathbb{Q}_{p}^{N}}}%
%BeginExpansion
{\displaystyle\int\limits_{\mathbb{Q}_{p}^{N}}}
%EndExpansion
\chi_{p}\left(  -\xi\cdot x\right)  \widehat{\Psi}_{0}(\xi)e^{-it||\xi
||_{p}^{\alpha}}d^{N}\xi\in\mathcal{C}(\mathbb{Q}_{p}^{N})\cap L^{2}%
(\mathbb{Q}_{p}^{N})\text{, for }t\geq0. \label{Solution}%
\end{equation}
Now considering $\Psi(\cdot,t)$, $t\in\mathbb{R}$, as a distribution from
$\mathcal{D}^{\prime}(\mathbb{Q}_{p}^{N})$, see the Appendix, and since
$e^{-it||\xi||_{p}^{\alpha}}\in L_{loc}^{1}(\mathbb{Q}_{p}^{N})$, we have
\[
\Psi(x,t)=\mathcal{F}_{\xi\rightarrow x}^{-1}(e^{-it||\xi||_{p}^{\alpha}}%
)\ast\Psi_{0}(x).
\]
Furthermore,
\[
\mathcal{F}_{\xi\rightarrow x}^{-1}(e^{-it||\xi||_{p}^{\alpha}})=Z(x,it)\text{
in }\mathcal{D}^{\prime}(\mathbb{Q}_{p}^{N})\text{, for }t\geq0,
\]
where $Z(x,t)$ is the heat kernel defined in (\ref{Heat_Kernel}). This last
equality follows from the fact that the Fourier transform is a homeomorphism
on $\mathcal{D}^{\prime}(\mathbb{Q}_{p}^{N})$. In conclusion, for $\Psi
_{0}(x)\in\mathcal{D}(\mathbb{Q}_{p}^{N})$,
\[
\Psi(x,t)=Z(x,it)\ast\Psi_{0}(x)=e^{-it\boldsymbol{H}}\Psi_{0}(x)\in
\mathcal{C}(\mathbb{Q}_{p}^{N})\cap L^{2}(\mathbb{Q}_{p}^{N}),
\]
for $t\geq0$.

\subsection{Born interpretation of the wave function}

Assume that
\begin{equation}%
%TCIMACRO{\dint \limits_{\mathbb{Q}_{p}^{N}}}%
%BeginExpansion
{\displaystyle\int\limits_{\mathbb{Q}_{p}^{N}}}
%EndExpansion
\left\vert \Psi_{0}(x)\right\vert ^{2}d^{N}x=1,
\label{Normalization_Condition}%
\end{equation}
and set $\varrho\left(  x,t\right)  :=\left\vert \Psi(x,t)\right\vert ^{2}$.
The interpretation of solution $\Psi(x,t)$ is that
\[%
%TCIMACRO{\dint \limits_{E}}%
%BeginExpansion
{\displaystyle\int\limits_{E}}
%EndExpansion
\varrho\left(  x,t\right)  d^{N}x
\]
is the probability of finding a single particle in the (Borel) subset
$E\subset\mathbb{Q}_{p}^{N}$ at the time $t$. Notice that by using the fact
that the Fourier transform preserves the inner product in $L^{2}%
(\mathbb{Q}_{p}^{N})$, we have
\begin{align*}%
%TCIMACRO{\dint \limits_{\mathbb{Q}_{p}^{N}}}%
%BeginExpansion
{\displaystyle\int\limits_{\mathbb{Q}_{p}^{N}}}
%EndExpansion
\varrho\left(  x,t\right)  d^{N}x  &  =%
%TCIMACRO{\dint \limits_{\mathbb{Q}_{p}^{N}}}%
%BeginExpansion
{\displaystyle\int\limits_{\mathbb{Q}_{p}^{N}}}
%EndExpansion
\Psi(x,t)\overline{\Psi(x,t)}d^{N}x=%
%TCIMACRO{\dint \limits_{\mathbb{Q}_{p}^{N}}}%
%BeginExpansion
{\displaystyle\int\limits_{\mathbb{Q}_{p}^{N}}}
%EndExpansion
\widehat{\Psi}(\xi,t)\overline{\widehat{\Psi}(\xi,t)}d^{N}\xi\\
&  =%
%TCIMACRO{\dint \limits_{\mathbb{Q}_{p}^{N}}}%
%BeginExpansion
{\displaystyle\int\limits_{\mathbb{Q}_{p}^{N}}}
%EndExpansion
\left\vert \widehat{\Psi_{0}}(\xi)\right\vert ^{2}d^{N}\xi=1.
\end{align*}

\section{\label{Section_7}de Broglie Wave-particle duality}

\subsection{The classical case}

1923 de Broglie postulates the existence of a correspondence between a beam of
free particles with momentum $\mathsf{p}$ and energy $\mathsf{E}$ with a wave
of form
\begin{equation}
\psi_{\mathbb{R}}\left(  x,t\right)  =Ae^{i\left(  \kappa\cdot x-\omega
t\right)  }\text{, \ \ \ }\kappa\in\mathbb{R}^{N}\text{, }\left(  x,t\right)
\in\mathbb{R}^{N+1}. \label{Plane_Waves}%
\end{equation}
A such correspondence should be relativistic so $\kappa\cdot x-\omega t$ is a
Lorentz invariant scalar product, which implies that the vectors $\left(
\mathsf{p},\mathsf{E}\right)  $ and $\left(  \kappa,\omega\right)  $ should be
proportional, see, e.g., \cite[pp 36-37]{Komech}. Now, by the Planck-Einsten
law $\mathsf{E}=\hbar\omega$, which implies that $\left(  \mathsf{p}%
,\mathsf{E}\right)  =\hbar\left(  \kappa,\omega\right)  $. The particle
wavelength is%
\[
\lambda=\frac{2\pi}{\left\vert \kappa\right\vert }=\frac{2\pi\hbar}{\left\vert
\mathsf{p}\right\vert }.
\]
Since the plane waves (\ref{Plane_Waves}) are solutions of the free
Schr\"{o}dinger equation; the de Broglie waves are `quantum waves.' We now
discuss these matters in the $p$-adic framework.

\subsection{The $p$-adic\ case}

Before discussing the de Broglie Wave-particle duality in the $p$-adic
framework, we require some preliminary results.

\subsubsection{Additive characters}

We recall that $p$-adic number $x\neq0$ has a unique expansion of the form
$x=p^{ord(x)}\sum_{j=0}^{\infty}x_{j}p^{j},$ where $x_{j}\in\{0,\dots,p-1\}$
and $x_{0}\neq0$. By using this expansion, we define the fractional part
of\textit{ }$x\in\mathbb{Q}_{p}$, denoted $\{x\}_{p}$, as the rational number
\[
\left\{  x\right\}  _{p}=\left\{
\begin{array}
[c]{lll}%
0 & \text{if} & x=0\text{ or }ord(x)\geq0\\
&  & \\
p^{ord(x)}\sum_{j=0}^{-ord(x)-1}x_{j}p^{j} & \text{if} & ord(x)<0.
\end{array}
\right.
\]
Set $\chi_{p}(y)=\exp(2\pi i\{y\}_{p})$ for $y\in\mathbb{Q}_{p}$. The map
$\chi_{p}(\cdot)$ is an additive character on $\mathbb{Q}_{p}$, i.e., a
continuous map from $\left(  \mathbb{Q}_{p},+\right)  $ into $S$ (the unit
circle considered as multiplicative group) satisfying $\chi_{p}(x_{0}%
+x_{1})=\chi_{p}(x_{0})\chi_{p}(x_{1})$, $x_{0},x_{1}\in\mathbb{Q}_{p}$. \ The
additive characters of $\mathbb{Q}_{p}$ form an Abelian group which is
isomorphic to $\left(  \mathbb{Q}_{p},+\right)  $. The isomorphism is given by
$\xi\rightarrow\chi_{p}(\xi x)$, see, e.g., \cite[Section 2.3]{Alberio et al}.

\subsubsection{\label{Section p-Adic Wavelets}$p$-Adic wavelets}

Given $\xi=\left(  \xi_{1},\ldots,\xi_{N}\right)  $, $x=\left(  x_{1}%
,\ldots,x_{N}\right)  \in\mathbb{Q}_{p}^{N}$, we set $\xi\cdot x=\sum
_{i=1}^{N}\xi_{i}x_{i}$. We denote by $\Omega\left(  p^{L}\left\Vert
x-a\right\Vert _{p}\right)  $ the characteristic function of the ball
\[
B_{-L}^{N}(a)=\left\{  x\in\mathbb{Q}_{p}^{N};\left\Vert x-a\right\Vert
_{p}\leq p^{-L}\right\}  =a+p^{L}\mathbb{Z}_{p}^{N}.
\]

We now define%
\[
\mathbb{Q}_{p}/\mathbb{Z}_{p}=\left\{  \sum_{j=-1}^{-m}x_{j}p^{j};\text{for
some }m>0\right\}  .
\]
For $b=\left(  b_{1},\ldots,b_{N}\right)  \in\left(  \mathbb{Q}_{p}%
/\mathbb{Z}_{p}\right)  ^{N}$, $r\in\mathbb{Z}$, we denote by $\Omega\left(
\left\Vert p^{r}x-b\right\Vert _{p}\right)  $ the characteristic function of
the ball $bp^{-r}+p^{-r}\mathbb{Z}_{p}^{N}$.

Set
\begin{equation}
\Psi_{rbk}\left(  x\right)  =p^{\frac{-rN}{2}}\chi_{p}(p^{-1}k\cdot\left(
p^{r}x-b\right)  )\Omega\left(  \left\Vert p^{r}x-b\right\Vert _{p}\right)  ,
\label{Basis_0}%
\end{equation}
where $r\in\mathbb{Z}$, $k=\left(  k_{1},\ldots,k_{N}\right)  \in
\{0,\dots,p-1\}^{N}$, $k\neq\left(  0,\ldots,0\right)  $, and $b=\left(
b_{1},\ldots,b_{N}\right)  \in\left(  \mathbb{Q}_{p}/\mathbb{Z}_{p}\right)
^{N}$. With this notation,%
\begin{equation}
\boldsymbol{D}^{\alpha}\Psi_{rbk}\left(  x\right)  =p^{\left(  1-r\right)
\alpha}\Psi_{rbk}\left(  x\right)  \text{, for any }r,n,k. \label{Basis}%
\end{equation}
Moreover,
\begin{equation}%
%TCIMACRO{\dint \limits_{\mathbb{Q}_{p}^{N}}}%
%BeginExpansion
{\displaystyle\int\limits_{\mathbb{Q}_{p}^{N}}}
%EndExpansion
\Psi_{rbk}\left(  x\right)  d^{N}x=0, \label{Average}%
\end{equation}
and $\left\{  \Psi_{rbk}\left(  x\right)  \right\}  _{rbk}$ forms a complete
orthonormal basis of $L^{2}(\mathbb{Q}_{p}^{N})$, see, e.g., \cite[Theorems
9.4.5 and 8.9.3]{Alberio et al}, \cite[Theorem 3.3]{KKZuniga}.

The wavelets $\Psi_{rbk}$ are obtained from the mother wavelet%
\[
\Psi_{k}\left(  x\right)  =\chi_{p}(p^{-1}k\cdot x)\Omega\left(  \left\Vert
x\right\Vert _{p}\right)
\]
as an orbit of the action of the group $x\rightarrow-n+p^{r}x$, $\Psi
_{rbk}\left(  x\right)  =p^{\frac{-Nr}{2}}\Psi_{k}\left(  -n+p^{r}x\right)  $,
see \cite{Alberio-Kozyrev}. The constant $p^{\frac{-Nr}{2}}$ is a
normalization factor so $\left\Vert \Psi_{rbk}\right\Vert _{2}=1$.

\subsubsection{$p$-Adic plane waves}

We now investigate the existence of plane waves for the free Schr\"{o}dinger
equation $i\frac{\partial\Psi(x,t)}{\partial t}=\boldsymbol{D}^{\alpha}%
\Psi(x,t)$. By separating variables as $\Psi(x,t)=\Psi_{time}(t)\Psi
_{space}(x)$ and using (\ref{Basis}), we have
\[
i\frac{\Psi_{time}^{\prime}(t)}{\Psi_{time}(t)}=\frac{\boldsymbol{D}^{\alpha
}\Psi_{space}(x)}{\Psi_{space}(x)}=\omega,
\]
and thus
\begin{align}
\Psi(x,t)  &  =e^{-ip^{\left(  1-r\right)  \alpha}t}\Psi_{rbk}\left(  x\right)
\nonumber\\
&  =A\exp\left\{  -i\left(  p^{\left(  1-r\right)  \alpha}t-2\pi\left\{
p^{r-1}k\cdot x\right\}  _{p}\right)  \right\}  \Omega\left(  \left\Vert
p^{r}x-b\right\Vert _{p}\right)  \text{, }\left(  x,t\right)  \in
\mathbb{Q}_{p}^{N}\times\mathbb{R}, \label{Plane_Waves_2}%
\end{align}
where $\omega=p^{\left(  1-r\right)  \alpha}$, and $A=A(k,b)\in\mathbb{C}$.
The quantity $p^{\left(  1-r\right)  \alpha}$ is an angular frequency, since
it is discrete, we postulate that the energy of wave (\ref{Plane_Waves_2}) is
\[
\mathsf{E}=\hbar p^{\left(  1-r\right)  \alpha}.
\]

\subsection{$p$-Adic de Broglie waves}

In standard QM the de Broglie waves are the plane wave solutions of the
Schr\"{o}dinger equation. By analogy, in the $p$-adic case, the de Broglie
waves have the form (\ref{Plane_Waves_2}). There are two central differences
between the $p$-adic case and the standard one. First, the scale group of
$\mathbb{Q}_{p}^{N}$ acts naturally on the plane waves (\ref{Plane_Waves_2})
such as we explained at the end of Section \ref{Section p-Adic Wavelets}.
Second,%
\begin{equation}
\text{all the }p\text{-adic planes waves (\ref{Plane_Waves_2}) have wavelength
}p^{-1}\text{.} \label{Wavelenght_Constancy}%
\end{equation}

This result implies that the de Broglie wave-particle duality does not hold in
the $p$-adic framework. It\ is an expected result since the smallest distance
in $\mathbb{Q}_{p}^{N}$ up to a scale transformation is $p^{-1}$. On the other
hand, To verify (\ref{Wavelenght_Constancy}), we first notice that the
wavelength of (\ref{Plane_Waves_2}) is determined by $\Psi_{rbk}\left(
x\right)  $, more precisely, by%
\[
\chi_{p}(p^{-1}k\cdot\left(  p^{r}x-b\right)  )\text{, for }x\in
p^{-r}b+p^{-r}\mathbb{Z}_{p}^{N}.
\]
By writing, $p^{r}x-b=j+p\mathbb{Z}_{p}^{N}$, where $j=j(x)\in\left\{
0,\ldots,p-1\right\}  ^{N}$, and using that the restriction of $\chi_{p}$ to
the unit ball is the constant function $1$, we have%
\begin{align}
\chi_{p}(p^{-1}k\cdot\left(  p^{r}x-b\right)  )  &  =\chi_{p}(p^{-1}k\cdot
j)\chi_{p}(k\cdot\mathbb{Z}_{p}^{N})=\chi_{p}(p^{-1}k\cdot j)=\chi_{p}(p^{-1}%
%TCIMACRO{\tsum \limits_{i=1}^{N}}%
%BeginExpansion
{\textstyle\sum\limits_{i=1}^{N}}
%EndExpansion
k_{i}j_{i})\nonumber\\
&  =\exp\left(  2\pi i\left\{  p^{-1}%
%TCIMACRO{\tsum \limits_{i=1}^{N}}%
%BeginExpansion
{\textstyle\sum\limits_{i=1}^{N}}
%EndExpansion
k_{i}j_{i}\right\}  _{p}\right)  =\exp\left(  i\frac{2\pi}{p}%
%TCIMACRO{\tsum \limits_{i=1}^{N}}%
%BeginExpansion
{\textstyle\sum\limits_{i=1}^{N}}
%EndExpansion
k_{i}j_{i}\right)  , \label{Wavelenght_Constancy_1}%
\end{align}
where in the last equality we use the convention that $k_{i}$, $j_{i}$ are
integers $\operatorname{mod}$ $p$, so the sum
\[%
%TCIMACRO{\tsum \limits_{i=1}^{N}}%
%BeginExpansion
{\textstyle\sum\limits_{i=1}^{N}}
%EndExpansion
k_{i}j_{i}\in\left\{  0,\ldots,p-1\right\}  .
\]
In this calculation $p\in\mathbb{Q}_{p}$, then the wavelength of
(\ref{Wavelenght_Constancy_1}) is $\left\vert p\right\vert _{p}=p^{-1}$.

By using the conclusions of the discussion about the motion in $\mathbb{Q}%
_{p}^{N}$ given at the end of Section \ref{Section_Motion}, we have
\begin{equation}
\text{the }p\text{-adic de Broglie waves are not physical but just
mathematical objects.} \label{DeBroglie-waves}%
\end{equation}

\section{The $p$-adic Schr\"{o}dinger equation with Gaussian initial data}

Our next goal is to study the two-slit experiment in the framework of the
$p$-adic QM. A key ingredient is a $p$-adic notion of Gaussian random
variables. Here, we use the notion of $\mathbb{Q}_{p}$-Gaussian $\mathbb{Q}%
_{p}$-valued random variables introduced by Evans, \cite{Evans}, see also
\cite[Chapter 6]{Kochubei}. A $p$-adic valued random variable $Y$ that is not
almost surely zero is $\mathbb{Q}_{p}$-Gaussian if and only if its
distribution is a cutoff of the Haar measure:%
\begin{equation}
\mathbb{P}(Y\in d^{N}x)=p^{-mN}\Omega\left(  p^{-m}\left\Vert x\right\Vert
_{p}\right)  d^{N}x \label{Probability measure}%
\end{equation}
for some $m\in\mathbb{Z}$, or equivalently%
\[
E(\chi_{p}\left(  \xi\cdot Y\right)  )=\Omega\left(  p^{m}\left\Vert
\xi\right\Vert _{p}\right)  \text{, \ }\xi\in\mathbb{Q}_{\mathfrak{p}}^{N},
\]
where $\Omega$ is the characteristic function of the interval $\left[
0,1\right]  $, see \cite[Theorem 6.1]{Kochubei}.

Notice that
\[
p^{-mN}\Omega\left(  p^{-m}\left\Vert x\right\Vert _{p}\right)  \text{
\ }\underrightarrow{\mathcal{F}_{x\rightarrow\xi}}\text{ \ }\Omega\left(
p^{m}\left\Vert \xi\right\Vert _{p}\right)  \text{, }m\in\mathbb{Z},
\]
in particular, $\mathcal{F}_{x\rightarrow\xi}\left(  \Omega\left(  \left\Vert
x\right\Vert _{p}\right)  \right)  =\Omega\left(  \left\Vert \xi\right\Vert
_{p}\right)  $. We interpret the probability measure
(\ref{Probability measure}) as a Gaussian localized particle.

\subsection{One-slit with Gaussian initial data}

\subsubsection{Some initial remarks}

We assume that at time $T=0$, without loss of generality, there is a Gaussian
localized particle confined in the ball $p^{L}\mathbb{Z}_{p}^{N}$, $L\geq0$.
We interpret the ball $p^{L}\mathbb{Z}_{p}^{N}$ as a small slit around the
origin. Following the classical formulation, see, e.g., \cite{McClendon and H.
Rabitz}-\cite{Webb-2} and the references therein, it is natural to propose the
following Gaussian initial datum:
\[
\Psi_{0}\left(  x\right)  =p^{\frac{LN}{2}}\Omega\left(  p^{L}\left\Vert
x\right\Vert _{p}\right)  \chi_{p}\left(  -k\cdot x\right)  \text{, }%
\]
where $L$ is a non-negative integer and $k\in\mathbb{Q}_{p}^{N}$. Notice that
$\Psi_{0}\left(  x\right)  \in\mathcal{D}(\mathbb{Q}_{p}^{N})$ satisfies
normalization condition (\ref{Normalization_Condition}). Furthermore, if $k\in
p^{-L}\mathbb{Z}_{p}^{N}$, i.e. $\left\Vert k\right\Vert _{p}\leq p^{L}$,
since $x\in p^{L}\mathbb{Z}_{p}^{N}$, i.e. $\left\Vert x\right\Vert _{p}\leq
p^{-L}$, $k\cdot x\in\mathbb{Z}_{p}$ and thus $\chi_{p}\left(  k\cdot
x\right)  \equiv1$. If $\left\Vert k\right\Vert _{p}>p^{L}$, then $\Psi
_{0}\left(  x\right)  $ is a character of the additive group $p^{L}%
\mathbb{Z}_{p}^{N}$, which is an eigenfunction of $\boldsymbol{D}^{\alpha}$:%
\[
\boldsymbol{D}^{\alpha}\Psi_{0}\left(  x\right)  =||k||_{p}^{\alpha}\Psi
_{0}\left(  x\right)  .
\]
This formula follows directly from the fact that%
\[
\widehat{\Psi}_{0}\left(  \xi\right)  =p^{\frac{-LN}{2}}\Omega\left(
p^{-L}\left\Vert \xi-k\right\Vert _{p}\right)  .
\]
The solution of Cauchy problem (\ref{Cauchy_Problem_1}) with initial datum
$\widehat{\Psi}_{0}\left(  \xi\right)  $ is%
\begin{align*}
\Psi\left(  x,t\right)   &  =p^{-\frac{LN}{2}}%
%TCIMACRO{\dint \limits_{\mathbb{Q}_{p}^{N}}}%
%BeginExpansion
{\displaystyle\int\limits_{\mathbb{Q}_{p}^{N}}}
%EndExpansion
\chi_{p}\left(  -\xi\cdot x\right)  e^{-it||\xi||_{p}^{\alpha}}\Omega\left(
p^{-L}\left\Vert \xi-k\right\Vert _{p}\right)  d^{N}\xi\\
&  =p^{\frac{-LN}{2}}%
%TCIMACRO{\dint \limits_{k+p^{-L}\mathbb{Z}_{p}^{N}}}%
%BeginExpansion
{\displaystyle\int\limits_{k+p^{-L}\mathbb{Z}_{p}^{N}}}
%EndExpansion
\chi_{p}\left(  -\xi\cdot x\right)  e^{-it||\xi||_{p}^{\alpha}}d^{N}\xi,
\end{align*}
see (\ref{Solution}). We now change variables as%
\[
\xi=k+p^{-L}y\text{, \ }d^{N}\xi=p^{LN}d^{N}y,
\]
and by the ultrametric property of the norm $||\cdot||_{p}$, $||k+p^{-L}%
y||_{p}^{\alpha}=||k||_{p}^{\alpha}$ for $\left\Vert k\right\Vert _{p}>p^{L}$
and any $\left\Vert y\right\Vert _{p}\leq1$, then%
\begin{align*}
\Psi\left(  x,t\right)   &  =p^{\frac{LN}{2}}\chi_{p}\left(  -x\cdot k\right)
%
%TCIMACRO{\dint \limits_{\mathbb{Z}_{p}^{N}}}%
%BeginExpansion
{\displaystyle\int\limits_{\mathbb{Z}_{p}^{N}}}
%EndExpansion
\chi_{p}\left(  -p^{-L}y\cdot x\right)  e^{-it||k+p^{-L}y||_{p}^{\alpha}}%
d^{N}y\\
&  =p^{\frac{LN}{2}}\chi_{p}\left(  -x\cdot k\right)  e^{-it||k||_{p}^{\alpha
}}%
%TCIMACRO{\dint \limits_{\mathbb{Z}_{p}^{N}}}%
%BeginExpansion
{\displaystyle\int\limits_{\mathbb{Z}_{p}^{N}}}
%EndExpansion
\chi_{p}\left(  -p^{-L}y\cdot x\right)  d^{N}y\\
&  =p^{\frac{LN}{2}}\chi_{p}\left(  -x\cdot k\right)  e^{-it||k||_{p}^{\alpha
}}\Omega\left(  \left\Vert p^{-L}x\right\Vert _{p}\right)  =e^{-it||k||_{p}%
^{\alpha}}\Psi_{0}\left(  x\right)  .
\end{align*}
Therefore,
\[
\left\vert \Psi\left(  x,t\right)  \right\vert ^{2}=\left\vert \Psi_{0}\left(
x\right)  \right\vert ^{2}.
\]
In this case the particle remains confined in the ball $p^{L}\mathbb{Z}%
_{p}^{N}$. Due to this reason, we do not use phase factors of the form
$\chi_{p}\left(  -k\cdot x\right)  $ in the initial datum.

\subsubsection{Single-slit diffraction}

Taking the considerations of the previous section, we set
\[
\Psi_{0}\left(  x\right)  =p^{\frac{LN}{2}}\Omega\left(  p^{L}\left\Vert
x\right\Vert _{p}\right)  ,
\]
where $L$ is a non-negative integer. The solution of Cauchy problem
(\ref{Cauchy_Problem_1}) with initial datum $\widehat{\Psi}_{0}\left(
\xi\right)  =p^{\frac{-LN}{2}}\Omega\left(  p^{-L}\left\Vert \xi\right\Vert
_{p}\right)  $ is%
\begin{align}
\Psi^{\left(  1-slit\right)  }\left(  x,t\right)   &  =p^{\frac{-LN}{2}}%
%TCIMACRO{\dint \limits_{\mathbb{Q}_{p}^{N}}}%
%BeginExpansion
{\displaystyle\int\limits_{\mathbb{Q}_{p}^{N}}}
%EndExpansion
\chi_{p}\left(  -\xi\cdot x\right)  e^{-it||\xi||_{p}^{\alpha}}\Omega\left(
p^{-L}\left\Vert \xi\right\Vert _{p}\right)  d^{N}\xi\nonumber\\
&  =p^{\frac{-LN}{2}}%
%TCIMACRO{\dint \limits_{p^{-L}\mathbb{Z}_{p}^{N}}}%
%BeginExpansion
{\displaystyle\int\limits_{p^{-L}\mathbb{Z}_{p}^{N}}}
%EndExpansion
\chi_{p}\left(  -\xi\cdot x\right)  e^{-it||\xi||_{p}^{\alpha}}d^{N}\xi.
\label{Solution_one_Slit}%
\end{align}

\paragraph{\textbf{Diffraction pattern near to the origin}}

For $\left\Vert x\right\Vert _{p}\leq p^{-L}$,%
\[
\left\vert \Psi^{\left(  1-slit\right)  }\left(  x,t\right)  \right\vert
^{2}\text{ is independent of }x\text{, and thus there is no a spatial
diffraction pattern.}%
\]
To establish this result, we first set
\[
\Psi_{0}^{\left(  1-slit\right)  }\left(  x,t\right)  =\Psi^{\left(
1-slit\right)  }\left(  x,t\right)  \Omega\left(  p^{L}\left\Vert x\right\Vert
_{p}\right)  .
\]
The condition $\left\Vert x\right\Vert _{p}\leq p^{-L}$ implies that
$x=p^{L}\widetilde{x}$, with $\left\Vert \widetilde{x}\right\Vert _{p}=1$.
Since $\xi=p^{-L}\widetilde{\xi}$, with $\left\Vert \widetilde{\xi}\right\Vert
_{p}=1$, we have $\xi\cdot x\in\mathbb{Z}_{p}$ thus $\chi_{p}\left(  -\xi\cdot
x\right)  =1$, and%
\[
\Psi_{0}^{\left(  1-slit\right)  }\left(  x,t\right)  =p^{\frac{-LN}{2}}%
%TCIMACRO{\dint \limits_{p^{-L}\mathbb{Z}_{p}^{N}}}%
%BeginExpansion
{\displaystyle\int\limits_{p^{-L}\mathbb{Z}_{p}^{N}}}
%EndExpansion
e^{-it||\xi||_{p}^{\alpha}}d^{N}\xi.
\]
By using the partition%
\begin{equation}
p^{-L}\mathbb{Z}_{p}^{N}=%
%TCIMACRO{\dbigsqcup \limits_{j=-L}^{\infty}}%
%BeginExpansion
{\displaystyle\bigsqcup\limits_{j=-L}^{\infty}}
%EndExpansion
p^{j}S_{0}^{N},\label{Partition}%
\end{equation}
where $S_{0}^{N}=\left\{  u\in\mathbb{Z}_{p}^{N};\left\Vert u\right\Vert
_{p}=1\right\}  $, we have%
\begin{align}
\Psi_{0}^{\left(  1-slit\right)  }\left(  x,t\right)   &  =p^{\frac{-LN}{2}}%
%TCIMACRO{\dsum \limits_{j=-L}^{\infty}}%
%BeginExpansion
{\displaystyle\sum\limits_{j=-L}^{\infty}}
%EndExpansion
\text{ }e^{-itp^{-j\alpha}}%
%TCIMACRO{\dint \limits_{||\xi||_{p}=p^{-j}}}%
%BeginExpansion
{\displaystyle\int\limits_{||\xi||_{p}=p^{-j}}}
%EndExpansion
d^{N}\xi=p^{\frac{-LN}{2}}\left(  1-p^{-N}\right)
%TCIMACRO{\dsum \limits_{j=-L}^{\infty}}%
%BeginExpansion
{\displaystyle\sum\limits_{j=-L}^{\infty}}
%EndExpansion
\text{ }e^{-itp^{-j\alpha}}p^{-jN}\nonumber\\
&  =p^{\frac{-LN}{2}}\left(  1-p^{-N}\right)  \left\{
%TCIMACRO{\dsum \limits_{j=-L}^{\infty}}%
%BeginExpansion
{\displaystyle\sum\limits_{j=-L}^{\infty}}
%EndExpansion
p^{-jN}\cos\left(  p^{-j\alpha}t\right)  +i%
%TCIMACRO{\dsum \limits_{j=-L}^{\infty}}%
%BeginExpansion
{\displaystyle\sum\limits_{j=-L}^{\infty}}
%EndExpansion
p^{-jN}\sin\left(  p^{-j\alpha}t\right)  \right\}  .\label{Psi-1-slit-0}%
\end{align}
Which shows that the diffraction pattern near the origin is independent of $x$.

\begin{remark}
By using that
\[
\left\vert
%TCIMACRO{\dsum \limits_{j=1}^{\infty}}%
%BeginExpansion
{\displaystyle\sum\limits_{j=1}^{\infty}}
%EndExpansion
p^{-jN}\cos\left(  p^{-j\alpha}t\right)  +i%
%TCIMACRO{\dsum \limits_{j=1}^{\infty}}%
%BeginExpansion
{\displaystyle\sum\limits_{j=1}^{\infty}}
%EndExpansion
p^{-jN}\sin\left(  p^{-j\alpha}t\right)  \right\vert \leq\frac{p^{-N}%
}{1-p^{-N}},
\]
for any $t\geq0$, one get the following approximation for $\Psi_{0}^{\left(
1-slit\right)  }\left(  x,t\right)  $:%
\[
\Psi_{0}^{\left(  1-slit\right)  }\left(  x,t\right)  \approx p^{\frac{-LN}%
{2}}\left(  1-p^{-N}\right)  \left\{
%TCIMACRO{\dsum \limits_{j=-L}^{0}}%
%BeginExpansion
{\displaystyle\sum\limits_{j=-L}^{0}}
%EndExpansion
p^{-jN}\cos\left(  p^{-j\alpha}t\right)  +i%
%TCIMACRO{\dsum \limits_{j=-L}^{0}}%
%BeginExpansion
{\displaystyle\sum\limits_{j=-L}^{0}}
%EndExpansion
p^{-jN}\sin\left(  p^{-j\alpha}t\right)  \right\}  ,
\]
for $p$ sufficiently large. Then%
\begin{align}
\left\vert \Psi_{0}^{\left(  1-slit\right)  }\left(  x,t\right)  \right\vert
^{2}  &  \approx p^{-LN}\left(  1-p^{-N}\right)  ^{2}\times\nonumber\\
&  \left\{  \left(
%TCIMACRO{\dsum \limits_{j=0}^{L}}%
%BeginExpansion
{\displaystyle\sum\limits_{j=0}^{L}}
%EndExpansion
p^{jN}\cos\left(  p^{j\alpha}t\right)  \right)  ^{2}+\left(
%TCIMACRO{\dsum \limits_{j=0}^{L}}%
%BeginExpansion
{\displaystyle\sum\limits_{j=0}^{L}}
%EndExpansion
p^{jN}\sin\left(  p^{j\alpha}t\right)  \right)  ^{2}\right\} \nonumber\\
&  =p^{-LN}\left(  1-p^{-N}\right)  ^{2}\left\{  L+1+%
%TCIMACRO{\dsum \limits_{j=0}^{L}}%
%BeginExpansion
{\displaystyle\sum\limits_{j=0}^{L}}
%EndExpansion
\text{ }%
%TCIMACRO{\dsum \limits_{0\leq j<k\leq L}}%
%BeginExpansion
{\displaystyle\sum\limits_{0\leq j<k\leq L}}
%EndExpansion
\text{ }p^{\left(  j+k\right)  N}\cos\left(  t\left(  p^{j\alpha}-p^{k\alpha
}\right)  \right)  \right\}  , \label{Approximation_Psi_1_slit_0}%
\end{align}
for $p$ sufficiently large.
\end{remark}

\paragraph{\textbf{Diffraction pattern at infinity}}

For $p$ sufficiently large, the diffraction pattern has the form
\begin{equation}
\left\vert \Psi^{\left(  1-slit\right)  }\left(  x,t\right)  \right\vert
^{2}\approx\frac{4p^{-LN}}{\left\Vert x\right\Vert _{p}^{2N}}\sin^{2}\left(
\frac{t\left(  1-p^{\alpha}\right)  }{2\left\Vert x\right\Vert _{p}^{\alpha}%
}\right)  \text{,} \label{Pattern_2}%
\end{equation}
for$\ t>0$, $\left\Vert x\right\Vert _{p}>p^{-L}$. Furthermore,%
\begin{equation}
\left\vert \Psi^{\left(  1-slit\right)  }\left(  x,t\right)  \right\vert
^{2}\leq\frac{4p^{-LN}}{\left\Vert x\right\Vert _{p}^{2N}}\text{ for }%
t\geq0\text{ and }\left\Vert x\right\Vert _{p}>p^{-L}\text{.} \label{Estimate}%
\end{equation}

This formula is established as follows. We set
\[
\Psi_{\infty}^{\left(  1-slit\right)  }\left(  x,t\right)  =\Psi\left(
x,t\right)  \left(  1-\Omega\left(  p^{L}\left\Vert x\right\Vert _{p}\right)
\right)  .
\]
The condition $\left\Vert x\right\Vert _{p}>p^{-L}$ implies that $-ord\left(
x\right)  >-L$. In this case, using (\ref{Solution_one_Slit}) and
(\ref{Partition}), and \cite[Chapter III, Lemma 4.1]{Taibleson}, we have%
\begin{align*}
\Psi_{\infty}^{\left(  1-slit\right)  }\left(  x,t\right)   &  =p^{\frac
{-LN}{2}}%
%TCIMACRO{\dsum \limits_{j=-L}^{\infty}}%
%BeginExpansion
{\displaystyle\sum\limits_{j=-L}^{\infty}}
%EndExpansion
\text{ }e^{-itp^{-j\alpha}}%
%TCIMACRO{\dint \limits_{||\xi||_{p}=p^{-j}}}%
%BeginExpansion
{\displaystyle\int\limits_{||\xi||_{p}=p^{-j}}}
%EndExpansion
\chi_{p}\left(  -\xi\cdot x\right)  d^{N}\xi\\
&  =p^{\frac{-LN}{2}}%
%TCIMACRO{\dsum \limits_{j=-L}^{\infty}}%
%BeginExpansion
{\displaystyle\sum\limits_{j=-L}^{\infty}}
%EndExpansion
\text{ }e^{-itp^{-j\alpha}}\left\{
\begin{array}
[c]{lll}%
p^{-jN}\left(  1-p^{-N}\right)  & \text{if } & \left\Vert x\right\Vert
_{p}\leq p^{j}\Leftrightarrow-ord\left(  x\right)  \leq j\\
&  & \\
-p^{-jN-N} & \text{if} & \left\Vert x\right\Vert _{p}=p^{j+1}\Leftrightarrow
-ord\left(  x\right)  =j+1\\
&  & \\
0 & \text{if} & \left\Vert x\right\Vert _{p}\geq p^{j+2}\Leftrightarrow
-ord\left(  x\right)  \geq j+2.
\end{array}
\right. \\
&  =p^{\frac{-LN}{2}}\left(  1-p^{-N}\right)
%TCIMACRO{\dsum \limits_{j=-ord\left(  x\right)  }^{\infty}}%
%BeginExpansion
{\displaystyle\sum\limits_{j=-ord\left(  x\right)  }^{\infty}}
%EndExpansion
p^{-jN}\text{ }e^{-itp^{-j\alpha}}-p^{\frac{-LN}{2}}p^{Nord\left(  x\right)
}e^{-itp^{\left(  ord\left(  x\right)  +1\right)  \alpha}}%
\end{align*}%
\begin{align}
&  =p^{\frac{-LN}{2}}\left(  1-p^{-N}\right)
%TCIMACRO{\dsum \limits_{j=-ord\left(  x\right)  }^{\infty}}%
%BeginExpansion
{\displaystyle\sum\limits_{j=-ord\left(  x\right)  }^{\infty}}
%EndExpansion
p^{-jN}\text{ }e^{-itp^{-j\alpha}}-p^{\frac{-LN}{2}}\left\Vert x\right\Vert
_{p}^{-N}e^{-it\left\Vert x\right\Vert _{p}^{-\alpha}p^{\alpha}}\nonumber\\
&  =p^{\frac{-LN}{2}}\left\Vert x\right\Vert _{p}^{-N}\left\{  \left(
1-p^{-N}\right)
%TCIMACRO{\dsum \limits_{m=0}^{\infty}}%
%BeginExpansion
{\displaystyle\sum\limits_{m=0}^{\infty}}
%EndExpansion
p^{-mN}\text{ }e^{-it\left\Vert x\right\Vert _{p}^{-\alpha}p^{-\alpha m}%
}-e^{-it\left\Vert x\right\Vert _{p}^{-\alpha}p^{\alpha}}\right\}  .
\label{Psi-1-slit-1}%
\end{align}
Estimate (\ref{Estimate}) follows from (\ref{Psi-1-slit-1}). By using that%
\[
\left\vert \left(  1-p^{-N}\right)
%TCIMACRO{\dsum \limits_{m=1}^{\infty}}%
%BeginExpansion
{\displaystyle\sum\limits_{m=1}^{\infty}}
%EndExpansion
p^{-mN}e^{-it\left\Vert x\right\Vert _{p}^{-\alpha}p^{-\alpha m}}\right\vert
\leq\left(  1-p^{-N}\right)
%TCIMACRO{\dsum \limits_{m=1}^{\infty}}%
%BeginExpansion
{\displaystyle\sum\limits_{m=1}^{\infty}}
%EndExpansion
p^{-mN}=p^{-N},
\]
for any $t\geq0$, $\left\Vert x\right\Vert _{p}>p^{-L}$, we have the following
approximation:%
\begin{equation}
\left(  1-p^{-N}\right)
%TCIMACRO{\dsum \limits_{m=0}^{\infty}}%
%BeginExpansion
{\displaystyle\sum\limits_{m=0}^{\infty}}
%EndExpansion
p^{-mN}\text{ }e^{-it\left\Vert x\right\Vert _{p}^{-\alpha}p^{-\alpha m}%
}\approx\left(  1-p^{-N}\right)  e^{-it\left\Vert x\right\Vert _{p}^{-\alpha}%
}\approx e^{-it\left\Vert x\right\Vert _{p}^{-\alpha}}, \label{Approximation}%
\end{equation}
for $p$ large. Therefore,%
\[
\Psi_{\infty}^{\left(  1-slit\right)  }\left(  x,t\right)  \approx
p^{\frac{-LN}{2}}\left\Vert x\right\Vert _{p}^{-N}\left(  e^{-it\left\Vert
x\right\Vert _{p}^{-\alpha}}-e^{-it\left\Vert x\right\Vert _{p}^{-\alpha
}p^{\alpha}}\right)  ,
\]
here, we used that $L\geq0$ to warranty that $p^{\frac{-LN}{2}}$ is very small
for $p$ large. Now
\begin{gather*}
\left\vert \Psi_{\infty}^{\left(  1-slit\right)  }\left(  x,t\right)
\right\vert ^{2}\approx2p^{-LN}\left\Vert x\right\Vert _{p}^{-2N}\times\\
\left(  1-\left\{  \cos\left(  t\left\Vert x\right\Vert _{p}^{-\alpha}\right)
\cos\left(  t\left\Vert x\right\Vert _{p}^{-\alpha}p^{\alpha}\right)
+\sin\left(  t\left\Vert x\right\Vert _{p}^{-\alpha}\right)  \sin\left(
t\left\Vert x\right\Vert _{p}^{-\alpha}p^{\alpha}\right)  \right\}  \right) \\
=2p^{-LN}\left\Vert x\right\Vert _{p}^{-2N}\left\{  1-\cos\left(  t\left\Vert
x\right\Vert _{p}^{-\alpha}\left(  1-p^{\alpha}\right)  \right)  \right\} \\
=4p^{-LN}\left\Vert x\right\Vert _{p}^{-2N}\sin^{2}\left(  t\left\Vert
x\right\Vert _{p}^{-\alpha}\left(  \frac{1-p^{\alpha}}{2}\right)  \right)  .
\end{gather*}

\subsection{Two-slit with Gaussian initial data}

We assume that at time zero, there is a localized particle $p^{\frac{LN}{2}%
}\Omega\left(  p^{L}\left\Vert x-a\right\Vert _{p}\right)  $ in a small slit
around a point $a$, and also there is another localized particle in a small
slit around a point $b$. Geometrically the first slit corresponds to the ball
$a+p^{L}\mathbb{Z}_{p}^{N}$ and the second one to the ball $b+p^{L}%
\mathbb{Z}_{p}^{N}$. We assume that the slits do not intersect. Based on these
facts, we use the following Gaussian initial datum:
\begin{equation}
\Psi_{0}\left(  x\right)  =\frac{1}{\sqrt{2}}p^{\frac{LN}{2}}\Omega\left(
p^{L}\left\Vert x-a\right\Vert _{p}\right)  +\frac{1}{\sqrt{2}}p^{\frac{LN}%
{2}}\Omega\left(  p^{L}\left\Vert x-b\right\Vert _{p}\right)  ,
\label{initial_datum_2_slit}%
\end{equation}
where $L$ is a non-negative integer, the points\ $a$, $b$ $\in\mathbb{Q}%
_{p}^{N}$ satisfy $\left\Vert a-b\right\Vert _{p}>p^{-L}$, which means that
the balls $a+p^{L}\mathbb{Z}_{p}^{N}$, $b+p^{L}\mathbb{Z}_{p}^{N}$ are
disjoint. The solution of Cauchy problem (\ref{Cauchy_Problem_1}) with initial
datum $\Psi_{0}\left(  x\right)  $ is%
\[
\Psi\left(  x,t\right)  =\Psi_{0,a}^{\left(  2-slit\right)  }\left(
x,t\right)  +\Psi_{\infty,a}^{\left(  2-slit\right)  }\left(  x,t\right)
+\Psi_{0,b}^{\left(  2-slit\right)  }\left(  x,t\right)  +\Psi_{\infty
,b}^{\left(  2-slit\right)  }\left(  x,t\right)  ,
\]
where
\[
\Psi_{0,a}^{\left(  2-slit\right)  }\left(  x,t\right)  =\frac{1}{\sqrt{2}%
}\Psi_{0}^{\left(  1-slit\right)  }\left(  x-a,t\right)  \text{, \ }\Psi
_{0,b}^{\left(  2-slit\right)  }\left(  x,t\right)  =\frac{1}{\sqrt{2}}%
\Psi_{0}^{\left(  1-slit\right)  }\left(  x-b,t\right)  ,
\]
see (\ref{Psi-1-slit-0}), and
\[
\Psi_{\infty,a}^{\left(  2-slit\right)  }\left(  x,t\right)  =\frac{1}%
{\sqrt{2}}\Psi_{\infty}^{\left(  1-slit\right)  }\left(  x-a,t\right)  \text{,
\ }\Psi_{\infty,b}^{\left(  2-slit\right)  }\left(  x,t\right)  =\frac
{1}{\sqrt{2}}\Psi_{\infty}^{\left(  1-slit\right)  }\left(  x-b,t\right)  ,
\]
see (\ref{Psi-1-slit-1}).

\textbf{Interference pattern at infinity}

We now consider points $x\in\mathbb{Q}_{p}^{N}$ satisfying
\[
\left\Vert x\right\Vert _{p}>\max\left\{  \left\Vert a\right\Vert
_{p},\left\Vert b\right\Vert _{p}\right\}  \text{ and }\left\Vert
x-a\right\Vert _{p}>p^{-L}\text{ and }\left\Vert x-b\right\Vert _{p}>p^{-L}.
\]
These points form an open neighborhood $\mathcal{N}_{\infty}$ of the infinity,
we denote by $\Psi_{\infty}^{\left(  2-slit\right)  }\left(  x,t\right)  $ the
restriction of $\Psi\left(  x,t\right)  $ to $\mathcal{N}_{\infty}$. Then%

\[
\left\vert \Psi_{\infty}^{\left(  2-slit\right)  }\left(  x,t\right)
\right\vert ^{2}\approx\frac{8p^{-LN}}{\sqrt{2}\left\Vert x\right\Vert
_{p}^{2N}}\sin^{2}\left(  \frac{t\left(  1-p^{\alpha}\right)  }{2\left\Vert
x\right\Vert _{p}^{\alpha}}\right)  ,\text{ for }x\in\mathcal{N}_{\infty},
\]
and $p$\ sufficiently large.

This formula is established as follows. The restriction $\Psi_{\infty
}^{\left(  2-slit\right)  }\left(  x,t\right)  $ of $\Psi\left(  x,t\right)  $
to $\mathcal{N}_{\infty}$ is given by
\begin{gather}
\Psi_{\infty}^{\left(  2-slit\right)  }\left(  x,t\right)  =\Psi_{\infty
}^{\left(  1-slit\right)  }\left(  x-a,t\right)  +\Psi_{\infty}^{\left(
1-slit\right)  }\left(  x-b,t\right) \nonumber\\
=\frac{1}{\sqrt{2}}p^{\frac{-LN}{2}}\left\Vert x-a\right\Vert _{p}%
^{-N}\left\{  \left(  1-p^{-N}\right)
%TCIMACRO{\dsum \limits_{m=0}^{\infty}}%
%BeginExpansion
{\displaystyle\sum\limits_{m=0}^{\infty}}
%EndExpansion
p^{-mN}\text{ }e^{-it\left\Vert x-a\right\Vert _{p}^{-\alpha}p^{-\alpha m}%
}-e^{-it\left\Vert x-a\right\Vert _{p}^{-\alpha}p^{\alpha}}\right\}
+\nonumber\\
\frac{1}{\sqrt{2}}p^{\frac{-LN}{2}}\left\Vert x-b\right\Vert _{p}^{-N}\left\{
\left(  1-p^{-N}\right)
%TCIMACRO{\dsum \limits_{m=0}^{\infty}}%
%BeginExpansion
{\displaystyle\sum\limits_{m=0}^{\infty}}
%EndExpansion
p^{-mN}\text{ }e^{-it\left\Vert x-b\right\Vert _{p}^{-\alpha}p^{-\alpha m}%
}-e^{-it\left\Vert x-b\right\Vert _{p}^{-\alpha}p^{\alpha}}\right\}  .
\label{Psi_2_slit_infinity}%
\end{gather}
In $\mathcal{N}_{\infty}$, the ultrametric property of $\left\Vert
\cdot\right\Vert _{p}$ implies that $\left\Vert x-a\right\Vert _{p}=\left\Vert
x-b\right\Vert _{p}=\left\Vert x\right\Vert _{p}$, then
(\ref{Psi_2_slit_infinity}) takes the form%
\[
\Psi_{\infty}^{\left(  2-slit\right)  }\left(  x,t\right)  =\frac{2}{\sqrt{2}%
}p^{\frac{-LN}{2}}\left\Vert x\right\Vert _{p}^{-N}\left\{  \left(
1-p^{-N}\right)
%TCIMACRO{\dsum \limits_{m=0}^{\infty}}%
%BeginExpansion
{\displaystyle\sum\limits_{m=0}^{\infty}}
%EndExpansion
p^{-mN}\text{ }e^{-it\left\Vert x\right\Vert _{p}^{-\alpha}p^{-\alpha m}%
}-e^{-it\left\Vert x\right\Vert _{p}^{-\alpha}p^{\alpha}}\right\}  .
\]
Now by using approximation (\ref{Approximation}), and reasoning as before, we
conclude that%
\begin{align}
\left\vert \Psi_{\infty}^{\left(  2-slit\right)  }\left(  x,t\right)
\right\vert ^{2}  &  \approx\left\vert \frac{2}{\sqrt{2}}p^{\frac{-LN}{2}%
}\left\Vert x\right\Vert _{p}^{-N}\left\{  \left(  1-p^{-N}\right)  \text{
}e^{-it\left\Vert x\right\Vert _{p}^{-\alpha}}-e^{-it\left\Vert x\right\Vert
_{p}^{-\alpha}p^{\alpha}}\right\}  \right\vert ^{2}\nonumber\\
&  =\frac{8}{\sqrt{2}}p^{-LN}\left\Vert x\right\Vert _{p}^{-2N}\sin^{2}\left(
t\left\Vert x\right\Vert _{p}^{-\alpha}\left(  \frac{1-p^{\alpha}}{2}\right)
\right)  , \label{Pattern_infinity}%
\end{align}
for $p$\ sufficiently large.

\textbf{Mid-range Interference pattern}

The set of mid-range points consist all the points $x\in\mathbb{Q}_{p}^{N}$
satisfying
\[
\left\Vert x\right\Vert _{p}\leq\max\left\{  \left\Vert a\right\Vert
_{p},\left\Vert b\right\Vert _{p}\right\}  \text{ and }\left\Vert
x-a\right\Vert _{p}>p^{-L}\text{ and }\left\Vert x-b\right\Vert _{p}>p^{-L}.
\]
We denote this set as $\mathcal{N}_{mid}$, and denote by\
\[
\Psi_{mid}^{\left(  2-slit\right)  }\left(  x,t\right)  =\Psi_{\infty
}^{\left(  1-slit\right)  }\left(  x-a,t\right)  +\Psi_{\infty}^{\left(
1-slit\right)  }\left(  x-b,t\right)  \text{, \ \ }x\in\mathcal{N}%
_{mid}\text{, }%
\]
\ the restriction of $\Psi\left(  x,t\right)  $ to $\mathcal{N}_{mid}$. The
description of $\left\vert \Psi_{mid}^{\left(  2-slit\right)  }\left(
x,t\right)  \right\vert ^{2}$ seems to be a difficult problem.

\textbf{Interference pattern near }$a$

The interference pattern near $a$ depends on the points $x\in\mathbb{Q}%
_{p}^{N}$ satisfying%
\[
\left\Vert x-a\right\Vert _{p}\leq p^{-L}\text{ and }\left\Vert x-b\right\Vert
_{p}>p^{-L}.
\]
We denote this set as $\mathcal{N}_{a}$, and denote by $\Psi_{a}^{\left(
2-slit\right)  }\left(  x,t\right)  $ the restriction of $\Psi\left(
x,t\right)  $ to $\mathcal{N}_{a}$, which is given by%
\[
\Psi_{a}^{\left(  2-slit\right)  }\left(  x,t\right)  =\frac{1}{\sqrt{2}}%
\Psi_{0}^{\left(  1-slit\right)  }\left(  x-a,t\right)  +\frac{1}{\sqrt{2}%
}\Psi_{\infty}^{\left(  1-slit\right)  }\left(  x-b,t\right)  ,
\]
for $x\in\mathcal{N}_{a}$. An approximation for $\left\vert \Psi_{a}^{\left(
2-slit\right)  }\left(  x,t\right)  \right\vert ^{2}$ is obtained from by
combining (\ref{Approximation_Psi_1_slit_0}) and (\ref{Pattern_2}). But, this
formula seems useful only for numerical calculations. The diffraction pattern
near $b$ is described in an analog way.

\subsection{The localized particles go only through one slit}

For the sake of simplicity we set $a=0$, and $b\in\mathbb{Q}_{p}^{N}$ with
$\left\Vert b\right\Vert _{p}>p^{-L}$ in (\ref{initial_datum_2_slit}).

The Fourier expansion of the localized particle $\Psi_{0}^{\left(  1\right)
}\left(  x\right)  =\frac{1}{\sqrt{2}}p^{\frac{LN}{2}}\Omega\left(
p^{L}\left\Vert x\right\Vert _{p}\right)  $ with respect to the basis
$\left\{  \Psi_{rbk}\left(  x\right)  \right\}  _{rbk}$ is given by%
\[
\Psi_{0}^{\left(  1\right)  }\left(  x\right)  =\frac{1}{\sqrt{2}}%
p^{-\frac{LN}{2}}%
%TCIMACRO{\dsum \limits_{r\geq-L+1}}%
%BeginExpansion
{\displaystyle\sum\limits_{r\geq-L+1}}
%EndExpansion%
%TCIMACRO{\dsum \limits_{k}}%
%BeginExpansion
{\displaystyle\sum\limits_{k}}
%EndExpansion
p^{-\frac{rN}{2}}\Psi_{r0k}\left(  x\right)  ,
\]
where the support of $\Psi_{r0k}\left(  x\right)  $ is $p^{-r}\mathbb{Z}%
_{p}^{N}$.

To establish this expansion, we first notice that the restriction of
$\Psi_{rbk}\left(  x\right)  $ to the ball $p^{L}\mathbb{Z}_{p}^{N}$ is given
by%
\begin{equation}
\Omega\left(  p^{L}\left\Vert x\right\Vert _{p}\right)  \Psi_{rbk}\left(
x\right)  =\left\{
\begin{array}
[c]{lll}%
\Psi_{rbk}\left(  x\right)  & \text{if} & bp^{-r}\in p^{L}\mathbb{Z}_{p}%
^{N}\text{, \ }r\leq-L\\
&  & \\
p^{-\frac{rN}{2}}\Omega\left(  p^{L}\left\Vert x\right\Vert _{p}\right)  &
\text{if} & bp^{-r}\in p^{-r}\mathbb{Z}_{p}^{N}\text{, \ }r\geq-L+1\\
&  & \\
0 & \text{if} & bp^{-r}\notin p^{-r}\mathbb{Z}_{p}^{N}\text{, \ }r\geq-L+1.
\end{array}
\right.  \label{Restriction}%
\end{equation}
The condition $bp^{-r}\in p^{-r}\mathbb{Z}_{p}^{N}$, \ $r\geq-L+1$ in the
second line in (\ref{Restriction}), is equivalent to $b=0$, and $r\geq-L+1$.

Now
\[
\Psi_{0}^{\left(  1\right)  }\left(  x\right)  =%
%TCIMACRO{\dsum \limits_{rnk}}%
%BeginExpansion
{\displaystyle\sum\limits_{rnk}}
%EndExpansion
C_{rnk}\Psi_{rnk}\left(  x\right)  \text{, with }C_{rnk}=%
%TCIMACRO{\dint \limits_{\mathbb{Q}_{p}^{N}}}%
%BeginExpansion
{\displaystyle\int\limits_{\mathbb{Q}_{p}^{N}}}
%EndExpansion
\Psi_{0}^{\left(  1\right)  }\left(  x\right)  \Psi_{rnk}\left(  x\right)
d^{N}x.
\]
If the support $\Psi_{rnk}\left(  x\right)  \subseteq p^{L}\mathbb{Z}_{p}^{N}$
(this case corresponds to the first line in (\ref{Restriction})), by
(\ref{Average}), $C_{rbk}=0$. If the support of $\Psi_{rbk}\left(  x\right)
\supsetneqq p^{L}\mathbb{Z}_{p}^{N}$ \ (this case corresponds \ to the second
line in (\ref{Restriction})), then
\[
C_{r0k}=\frac{1}{\sqrt{2}}p^{\frac{LN}{2}}%
%TCIMACRO{\dint \limits_{\mathbb{Q}_{p}^{N}}}%
%BeginExpansion
{\displaystyle\int\limits_{\mathbb{Q}_{p}^{N}}}
%EndExpansion
p^{-\frac{rN}{2}}\Omega\left(  p^{L}\left\Vert x\right\Vert _{p}\right)
d^{N}x=\frac{1}{\sqrt{2}}p^{\frac{-LN}{2}-\frac{rN}{2}}\text{, }r\geq-L+1.
\]
Therefore,
\[
\Psi_{0}^{\left(  1\right)  }\left(  x\right)  =%
%TCIMACRO{\dsum \limits_{r\geq-L+1}}%
%BeginExpansion
{\displaystyle\sum\limits_{r\geq-L+1}}
%EndExpansion%
%TCIMACRO{\dsum \limits_{k}}%
%BeginExpansion
{\displaystyle\sum\limits_{k}}
%EndExpansion
\frac{1}{\sqrt{2}}p^{\frac{-LN}{2}-\frac{rN}{2}}\Psi_{r0k}\left(  x\right)  .
\]
Now the particle $\Psi_{0}^{\left(  1\right)  }\left(  x\right)  $ is a
superposition of the `$p$-adic waves' $\Psi_{r0k}\left(  x\right)  $,
$r\geq-L+1$, $k\in\{0,\dots,p-1\}^{N}$, $\kappa\neq\left(  0,\ldots,0\right)
$. Notice that for or $r$ sufficiently large, the support of $\Psi
_{r0k}\left(  x\right)  $ contains point $b$.%
\begin{equation}
\text{If the }\Psi_{r0k}\left(  x\right)  \text{ can be interpreted as
oscillanting waves in }\mathbb{Q}_{p}^{N}, \label{condition}%
\end{equation}
most of these waves, i.e., for $r$ sufficiently large, will `interfere' with
the components of the particle $\frac{1}{\sqrt{2}}p^{\frac{LN}{2}}%
\Omega\left(  p^{L}\left\Vert x-b\right\Vert _{p}\right)  $ located around
$b$. This implies that the localized particle $\Psi_{0}^{\left(  1\right)
}\left(  x\right)  $ goes through both slits. Considering that we already
argued (\ref{DeBroglie-waves}), that the de Broglie waves are just
mathematical objects and that \ the notion of trajectory $t\rightarrow
X(t)\in$ $\mathbb{Q}_{p}^{N}$ rules out the possibility that a particle is at
the same in two different points, we argue that each particle $\frac{1}%
{\sqrt{2}}p^{\frac{LN}{2}}\Omega\left(  p^{L}\left\Vert x-a\right\Vert
_{p}\right)  $, $\frac{1}{\sqrt{2}}p^{\frac{LN}{2}}\Omega\left(
p^{L}\left\Vert x-b\right\Vert _{p}\right)  $ goes only through one slit.

\section{Discussion}

The Bronstein inequality asserts that the uncertainty of any length
measurement is greater than the Planck length. By interpreting this inequality
as the nonexistence of `intervals' below the Planck scale, we interpret this
fact as the physical space at a very short distance is a completely
disconnected topological space; intuitively, the space is just a collection of
isolated points. Furthermore, given two different points $a$, $b$ in this
space, there are no continuous trajectories, $X(t)$, $t\in\mathbb{R}_{+}$,
satisfying $X(t_{1})=a$, $X(t_{2})=b$. In turn, this fact implies that motion
in a completely disconnected space is just a sequence of jumps between the
points of the space. This article assumes that the physical space is discrete
and uses $\mathbb{Q}_{p}^{N}$ as a model. In principle, $N=3$, but it is
convenient to formulate our results in arbitrary dimensions considering other
applications. Using $\mathbb{Q}_{p}^{N}$ as a model of the physical space at
short distances, our notion of discreteness is just the Volovich conjecture on
the $p$-adic nature of the space below the Planck length. In this framework,
$p^{-1}$ is the smallest distance between two different points, up to scale transformations.

$p$-Adic QM is constructed from the Dirac-von Neumann axioms, assuming that
the quantum states vectors from $L^{2}(\mathbb{Q}_{p}^{N})$. The evolution
operators have the form $e^{-it\boldsymbol{D}^{\alpha}}$, $\alpha>0$, for
$t\in\mathbb{R}$; they are obtained from a Feller semigroup
$e^{-t\boldsymbol{D}^{\alpha}}$, $t\in\mathbb{R}_{+}$, by a Wick rotation. The
Feller feature means that there is a Markov process attached to the semigroup
with state space $\mathbb{Q}_{p}^{N}$, with discontinuous paths. Intuitively,
this stochastic process describes a particle performing a random motion
consisting of jumps between the points of a self-similar set.

The free Schr\"{o}dinger equation in natural units has the form%
\begin{equation}
\left\{
\begin{array}
[c]{ll}%
i\frac{\partial\Psi\left(  x,t\right)  }{\partial t}=\boldsymbol{D}^{\alpha
}\Psi\left(  x,t\right)  , & x\in\mathbb{Q}_{p}^{N},t\geq0\\
& \\
\Psi\left(  x,0\right)  =\Psi_{0}\left(  x\right)  . &
\end{array}
\right.  \label{Equation_11}%
\end{equation}
The operator $\boldsymbol{D}^{\alpha}$ is nonlocal, i.e., $\left(
\boldsymbol{D}^{\alpha}\phi\right)  \left(  x_{0}\right)  $ depends on all
points of $\mathbb{Q}_{p}^{N}$. The function $\left\vert \Psi\left(
x,t\right)  \right\vert ^{2}$ is a time-dependent probability density. In
particular, $\int_{B}\left\vert \Psi\left(  x,t\right)  \right\vert ^{2}%
d^{N}x$ is the probability of finding a particle in the set $B$ at the time
$t$. The particle interpretation comes from the fact that the Schr\"{o}dinger
equation comes from a heat equation by a Wick rotation and that this last
equation describes a particle jumping randomly in $\mathbb{Q}_{p}^{N}$.

This equation admits plane wave solutions of the form%
\begin{equation}
\varkappa_{p}\left(  x,t\right)  =\exp\left(  -i\left(  p^{\left(  1-r\right)
\alpha}t-2\pi\left\{  p^{r-1}k\cdot x\right\}  _{p}\right)  \right)
\Omega\left(  \left\Vert p^{r}x-n\right\Vert _{p}\right)  \text{, }\left(
x,t\right)  \in\mathbb{Q}_{p}^{N}\times\mathbb{R}\text{,} \label{Beam}%
\end{equation}
where $r\in\mathbb{Z}$, and $k$, $n$ are suitable vectors from $\mathbb{Q}%
_{p}^{N}$. By analogy with the standard QM, we call all these waves the de
Broglie waves. The scale group of $\mathbb{Q}_{p}^{N}$ acts on the
$\varkappa_{p}\left(  x,t\right)  $, and all these `waves' have the same
wavelength $p^{-1}$. Here, we mention that this fact is consistent with a
`wave' propagation on a lattice where a point is a distance $p^{-1}$ to its
neighbors. Thus, in the $p$-adic context, the wave-particle duality ceases to
be valid. We interpret this fact as that the standard wave-particle duality is
a manifestation of the discreteness of the space.

We argue that the `waves' $\varkappa_{p}\left(  x,t\right)  \in\mathbb{C}$ do
not have physical meaning; more precisely, the `wave' $\varkappa_{p}\left(
x,t\right)  $ does not describe an oscillation in $\mathbb{Q}_{p}^{N}$. Since
any wavefunction in $L^{2}(\mathbb{Q}_{p}^{N})$ can be represented as a series
in the `waves' $\varkappa_{p}\left(  x,t\right)  $, the wavefunctions
themselves do not have physical meaning, only the probability densities that
they define have it. In the $p$-adic framework, the probability densities
$\left\vert \Psi\left(  x,t\right)  \right\vert ^{2}$ exhibit `interference
patterns' similar to the standard ones. This suggests that the $p$-adic
Schr\"{o}dinger equations provide pictures\ similar to the standard ones.
However, the exact comparison of the $p$-adic interference patterns with the
classical ones is an open problem.

Our main result is a model of the double-slit experiment in QM. In our
description, at time zero, there are two localized particles, which correspond
to two localized Gaussian \ probability densities at two different points $a$,
$b\in\mathbb{Q}_{p}^{N}$, such $\left\Vert a-b\right\Vert >p^{-L}$.
Geometrically, each slit corresponds to a ball centered at $a$ or $b$ with
radius $p^{-L}$, $L\geq0$, which is the size of each slit. These localized
particles evolve in time under a nonlocal evolution operator
$e^{-it\boldsymbol{D}^{\alpha}}$, $t\geq0$; the nonlocality is a consequence
of the discreteness of space $\mathbb{Q}_{p}^{N}$. In standard QM, the fact
that the de Broglie waves (the quantum waves) are considered physical waves
explains the double-slit experiment as an interference phenomenon.

In the $p$-adic framework, de Broglie waves (`the quantum waves') are not
physical entities; thus, the double-slit experiment explanation as an
interference phenomenon is ruled out. But, the probability density $\left\vert
\Psi\left(  x,t\right)  \right\vert ^{2}$ exhibits the classical interference
patterns attributed to `quantum waves.' For this reason, we\ call $\left\vert
\Psi\left(  x,t\right)  \right\vert ^{2}$\ a diffraction/interference pattern.
We argue that the standard notion of trajectory, jointly with the fact that
the de Broglie waves are just mathematical objects, concludes that in the
double-slit experiment, each particle goes only through one of the slits. It
is relevant to mention that the same conclusion was posed in \cite{Aharonov et
al}; in this article, the authors proposed nonlocal interactions between the
slits as the key to understanding the double-slit experiment. In our
framework, the nonlocal interactions result from the discreteness of space
$\mathbb{Q}_{p}^{N}$.

We study interference diffraction patterns $\left\vert \Psi\left(  x,t\right)
\right\vert ^{2}$ for the cases of two slits and one slit. In the case one
slit located in a small ball around the origin, the diffraction pattern
$\left\vert \Psi\left(  x,t\right)  \right\vert ^{2}$ consists of two regions.
Near to the origin $\left\vert \Psi\left(  x,t\right)  \right\vert ^{2}$ is a
function depending only on $t$, which means that there is no spatial
diffraction. If $p$\ is sufficiently large, by taking $\left\Vert x\right\Vert
_{p}=p^{-L+1}$ in (\ref{Pattern_infinity}), which is equivalent to observe the
interference pattern on a screen at distance $p^{-L+1}$ from the origin, we
have
\[
\left\vert \Psi\left(  p^{L-1},t\right)  \right\vert ^{2}\approx\frac{8p^{-N}%
}{\sqrt{2}}\mathit{\sin}^{2}\left(  \frac{p^{L\alpha}t}{2}\right)  \text{, for
}t\geq0\text{.}%
\]
Notice that $p^{L\alpha}$ is an eigenvalue of $\boldsymbol{D}^{\alpha}$.

The interference pattern of the double-slit experiment consists of four
different regions. If $\left\Vert x\right\Vert _{p}$ is sufficiently large,
the interference pattern looks like one of one-slit. A mid-range interference
pattern depends on both slits, and there are two additional patterns around
each slit. All these patterns show interference in space and time. The
interference patterns rely heavily on the symbol of operator $\boldsymbol{D}%
^{\alpha}$.

\begin{figure}[ptb]
\begin{center}
\includegraphics[height=2.76in,width=3.98in,angle=0]{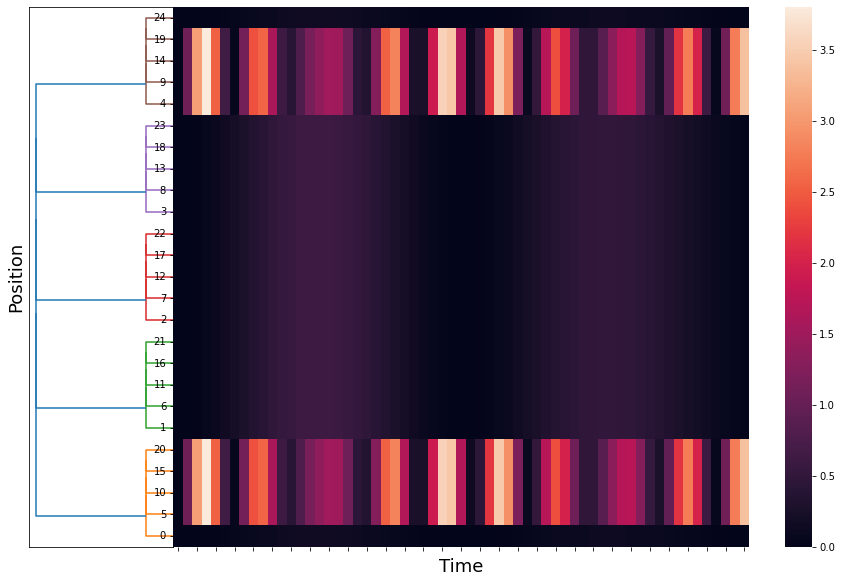}
\end{center}
\caption{ Numerical approximation for the interference pattern $\left\vert
\Psi\left( x,t\right) \right\vert ^{2}$ produced by a two slit experiment in
the mid-range zone in dimension $1$. Take $p=5$, $L=2$, $a=0$ ($\left\vert
a\right\vert _{5}=0$), $b=24$ ($\left\vert b\right\vert _{5}=1$). The
mid-range zone is then $\left\vert x\right\vert _{5}\leq1$ (the unit \ ball $%
\mathbb{Z}_{5}$). The unit ball is approximated by the finite tree $G_{2}=%
\mathbb{Z}_{5}/5^{2}\mathbb{Z}_{5}$. The elements of this tree constitute
the vertical scale. Horizontal scale corresponds to the time running on the
interval $\left[ 0,3\right] $, with a step of $0.05$.}
\label{Figure 3}
\end{figure}

\begin{figure}[ptb]
\begin{center}
\includegraphics[height=2.00in,width=5.73in,angle=0]{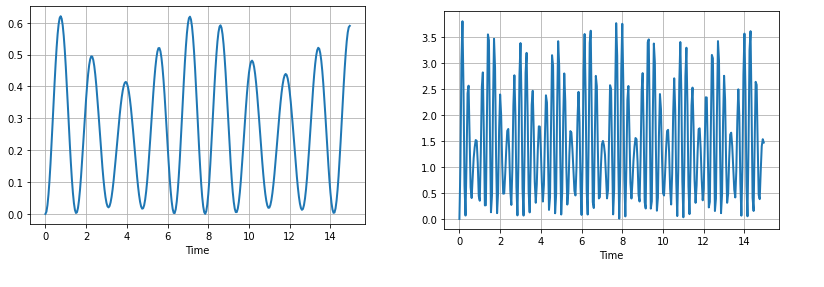}
\end{center}
\caption{The left figure shows the interference pattern $\left\vert
\Psi\left( 6,t\right) \right\vert ^{2}$, for $x=6$ ($\left\vert 6\right\vert
_{5}=1$) and $t$ running in the interval $\left[ 0,15\right] $. The right
figure shows the interference pattern $\left\vert \Psi\left( 5,t\right)
\right\vert ^{2}$, for $x=5$ ($\left\vert 5\right\vert _{5}=5^{-1}$) and $t$
running in the interval $\left[ 0,15\right] $. }
\label{Figure 4}
\end{figure}

In dimension $N=1$, by using formula (\ref{Psi-1-slit-1}), with \ $25$ terms,
$p=5$, $L=2$, \ we construct a numerical approximation for $\left\vert
\Psi\left(  x,t\right)  \right\vert ^{2}$ in the mid-range zone, see Figure
\ref{Figure 3}. By fixing specific points, Figure \ref{Figure 4} shows the
interference patterns $\left\vert \Psi\left(  6,t\right)  \right\vert ^{2}$,
$\left\vert \Psi\left(  5,t\right)  \right\vert ^{2}$. These numerical
simulations aim not to predict any specific interference pattern but to show
that the obtained interference patterns have the expected characteristics.%

\section{\label{Appendix} Appendix: basic facts on $p$-adic analysis}

In this section, we fix the notation and collect some basic results on $p$%
-adic analysis that we will use throughout the article. For a detailed
exposition on $p$-adic analysis the reader may consult \cite{V-V-Z}, \cite%
{Alberio et al}-\cite{Taibleson}.

\subsection{The field of $p$-adic numbers}

In this article, $p$ denotes a prime number. The field of $p-$adic numbers 
$\mathbb{Q}_{p}$ is defined as the completion of the field of rational
numbers $\mathbb{Q}$ with respect to the $p-$adic norm $|\cdot|_{p}$, which
is defined as 
\begin{equation*}
|x|_{p}=%
\begin{cases}
0 & \text{if }x=0 \\ 
p^{-\gamma} & \text{if }x=p^{\gamma}\dfrac{a}{b},%
\end{cases}%
\end{equation*}
where $a$ and $b$ are integers coprime with $p$. The integer $\gamma
=ord_{p}(x):=ord(x)$, with $ord(0):=+\infty$, is called the\textit{\ }$p-$%
\textit{adic order of} $x$. We extend the $p-$adic norm to $\mathbb{Q}%
_{p}^{N}$ by taking%
\begin{equation*}
||x||_{p}:=\max_{1\leq i\leq N}|x_{i}|_{p},\qquad\text{for }%
x=(x_{1},\dots,x_{N})\in\mathbb{Q}_{p}^{N}.
\end{equation*}
By defining $ord(x)=\min_{1\leq i\leq N}\{ord(x_{i})\}$, we have $%
||x||_{p}=p^{-ord(x)}$.\ The metric space $\left( \mathbb{Q}%
_{p}^{N},||\cdot||_{p}\right) $ is a complete ultrametric space. As a
topological space $\mathbb{Q}_{p}$\ is homeomorphic to a Cantor-like subset
of the real line, see, e.g., \cite{V-V-Z}, \cite{Alberio et al}.

Any $p-$adic number $x\neq0$ has a unique expansion of the form 
\begin{equation*}
x=p^{ord(x)}\sum_{j=0}^{\infty}x_{j}p^{j},
\end{equation*}
where $x_{j}\in\{0,1,2,\dots,p-1\}$ and $x_{0}\neq0$. \ In addition, any $%
x\in\mathbb{Q}_{p}^{N}\smallsetminus\left\{ 0\right\} $ can be represented
uniquely as $x=p^{ord(x)}v$, where $\left\Vert v\right\Vert _{p}=1$.

\subsection{Topology of $\mathbb{Q}_{p}^{N}$}

For $r\in\mathbb{Z}$, denote by $B_{r}^{N}(a)=\{x\in\mathbb{Q}%
_{p}^{N};||x-a||_{p}\leq p^{r}\}$ the ball of radius $p^{r}$ with center at $%
a=(a_{1},\dots,a_{N})\in\mathbb{Q}_{p}^{N}$, and take $%
B_{r}^{N}(0):=B_{r}^{N}$. Note that $B_{r}^{N}(a)=B_{r}(a_{1})\times\cdots%
\times B_{r}(a_{N})$, where $B_{r}(a_{i}):=\{x\in\mathbb{Q}%
_{p};|x_{i}-a_{i}|_{p}\leq p^{r}\}$ is the one-dimensional ball of radius $%
p^{r}$ with center at $a_{i}\in\mathbb{Q}_{p}$. The ball $B_{0}^{N}$ equals
the product of $N$ copies of $B_{0}=\mathbb{Z}_{p}$, the ring of $p-$adic
integers. We also denote by $S_{r}^{N}(a)=\{x\in\mathbb{Q}%
_{p}^{N};||x-a||_{p}=p^{r}\}$ the sphere of radius $p^{r}$ with center at $%
a=(a_{1},\dots,a_{N})\in \mathbb{Q}_{p}^{N}$, and take $%
S_{r}^{N}(0):=S_{r}^{N}$. We notice that $S_{0}^{1}=\mathbb{Z}_{p}^{\times}$
(the group of units of $\mathbb{Z}_{p}$), but $\left( \mathbb{Z}%
_{p}^{\times}\right) ^{N}\subsetneq S_{0}^{N}$. The balls and spheres are
both open and closed subsets in $\mathbb{Q}_{p}^{N}$. In addition, two balls
in $\mathbb{Q}_{p}^{N}$ are either disjoint or one is contained in the other.

As a topological space $\left( \mathbb{Q}_{p}^{N},||\cdot||_{p}\right) $ is
totally disconnected, i.e., the only connected \ subsets of $\mathbb{Q}%
_{p}^{N}$ are the empty set and the points. A subset of $\mathbb{Q}_{p}^{N}$
is compact if and only if it is closed and bounded in $\mathbb{Q}_{p}^{N}$,
see, e.g., \cite[Section 1.3]{V-V-Z}, or \cite[Section 1.8]{Alberio et al}.
The balls and spheres are compact subsets. Thus $\left( \mathbb{Q}%
_{p}^{N},||\cdot||_{p}\right) $ is a locally compact topological space.

\subsection{The Haar measure}

Since $(\mathbb{Q}_{p}^{N},+)$ is a locally compact topological group, there
exists a Haar measure $d^{N}x$, which is invariant under translations, i.e., 
$d^{N}(x+a)=d^{N}x$, \cite{Halmos}. If we normalize this measure by the
condition $\int_{\mathbb{Z}_{p}^{N}}dx=1$, then $d^{N}x$ is unique.

\begin{notation}
We will use $\Omega\left( p^{-r}||x-a||_{p}\right) $ to denote the
characteristic function of the ball $B_{r}^{N}(a)=a+p^{-r}\mathbb{Z}_{p}^{N}$%
, where 
\begin{equation*}
\mathbb{Z}_{p}^{N}=\left\{ x\in\mathbb{Q}_{p}^{N};\left\Vert x\right\Vert
_{p}\leq1\right\}
\end{equation*}
is the $N$-dimensional unit ball. For more general sets, we will use the
notation $1_{A}$ for the characteristic function of set $A$.
\end{notation}

\subsection{The Bruhat-Schwartz space}

A complex-valued function $\varphi$ defined on $\mathbb{Q}_{p}^{N}$ is
called locally constant if for any $x\in\mathbb{Q}_{p}^{N}$ there exist an
integer $l(x)\in\mathbb{Z}$ such that%
\begin{equation}
\varphi(x+x^{\prime})=\varphi(x)\text{ for any }x^{\prime}\in B_{l(x)}^{N}.
\label{local_constancy}
\end{equation}
A function $\varphi:\mathbb{Q}_{p}^{N}\rightarrow\mathbb{C}$ is called a
Bruhat-Schwartz function (or a test function) if it is locally constant with
compact support. Any test function can be represented as a linear
combination, with complex coefficients, of characteristic functions of
balls. The $\mathbb{C}$-vector space of Bruhat-Schwartz functions is denoted
by $\mathcal{D}(\mathbb{Q}_{p}^{N})$. For $\varphi\in\mathcal{D}(\mathbb{Q}%
_{p}^{N})$, the largest number $l=l(\varphi)$ satisfying (\ref%
{local_constancy}) is called the exponent of local constancy (or the
parameter of constancy) of $\varphi$.

We denote by $\mathcal{D}_{m}^{l}(\mathbb{Q}_{p}^{N})$ the
finite-dimensional space of test functions from $\mathcal{D}(\mathbb{Q}%
_{p}^{N})$ having supports in the ball $B_{m}^{N}$ and with parameters \ of
constancy $\geq l$. We now define a topology on $\mathcal{D}(\mathbb{Q}%
_{p}^{N})$ as follows. We say that a sequence $\left\{ \varphi_{j}\right\}
_{j\in\mathbb{N}}$ of functions in $\mathcal{D}(\mathbb{Q}_{p}^{N})$
converges to zero, if the two following conditions hold true:

(1) there are two fixed integers $k_{0}$ and $m_{0}$ such that \ each $%
\varphi_{j}\in$ $\mathcal{D}_{m_{0}}^{k_{0}}(\mathbb{Q}_{p}^{N})$;

(2) $\varphi_{j}\rightarrow0$ uniformly.

$\mathcal{D}(\mathbb{Q}_{p}^{N})$ endowed with the above topology becomes a
topological vector space.

\subsection{$L^{\protect\rho}$ spaces}

Given $\rho\in\lbrack0,\infty)$, we denote by$L^{\rho}\left( 
%TCIMACRO{\U{211a} }%
%BeginExpansion
\mathbb{Q}
%EndExpansion
_{p}^{N}\right) :=L^{\rho}\left( 
%TCIMACRO{\U{211a} }%
%BeginExpansion
\mathbb{Q}
%EndExpansion
_{p}^{N},d^{N}x\right) ,$ the $\mathbb{C}-$vector space of all the complex
valued functions $g$ satisfying 
\begin{equation*}
\left\Vert g\right\Vert _{\rho}=\left( \text{ }\dint \limits_{\mathbb{Q}%
_{p}^{N}}\left\vert g\left( x\right) \right\vert ^{\rho}d^{N}x\right) ^{%
\frac {1}{\rho}}<\infty,
\end{equation*}
where $d^{N}x$ is the normalized Haar measure on $\left( \mathbb{Q}%
_{p}^{N},+\right) $.

If $U$ is an open subset of $\mathbb{Q}_{p}^{N}$, $\mathcal{D}(U)$ denotes
the $\mathbb{C}$-vector space of test functions with supports contained in $%
U $, then $\mathcal{D}(U)$ is dense in 
\begin{equation*}
L^{\rho}\left( U\right) =\left\{ \varphi:U\rightarrow\mathbb{C};\left\Vert
\varphi\right\Vert _{\rho}=\left\{ \dint \limits_{U}\left\vert \varphi\left(
x\right) \right\vert ^{\rho}d^{N}x\right\} ^{\frac{1}{\rho}}<\infty\right\} ,
\end{equation*}
for $1\leq\rho<\infty$, see, e.g., \cite[Section 4.3]{Alberio et al}. We
denote by $L_{\mathbb{R}}^{\rho}\left( U\right) $ the real counterpart of $%
L^{\rho}\left( U\right) $.

\subsection{The Fourier transform}

Set $\chi_{p}(y)=\exp(2\pi i\{y\}_{p})$ for $y\in\mathbb{Q}_{p}$, and $%
\xi\cdot x:=\sum_{j=1}^{N}\xi_{j}x_{j}$, for $\xi=(\xi_{1},\dots,\xi_{N})$, $%
x=(x_{1},\dots,x_{N})\allowbreak\in\mathbb{Q}_{p}^{N}$, as before. The
Fourier transform of $\varphi\in\mathcal{D}(\mathbb{Q}_{p}^{N})$ is defined
as 
\begin{equation*}
\mathcal{F}\varphi(\xi)=\dint \limits_{\mathbb{Q}_{p}^{N}}\chi_{p}(\xi\cdot
x)\varphi(x)d^{N}x\quad\text{for }\xi\in\mathbb{Q}_{p}^{N},
\end{equation*}
where $d^{N}x$ is the normalized Haar measure on $\mathbb{Q}_{p}^{N}$. The
Fourier transform is a linear isomorphism from $\mathcal{D}(\mathbb{Q}%
_{p}^{N})$ onto itself satisfying 
\begin{equation}
(\mathcal{F}(\mathcal{F}\varphi))(\xi)=\varphi(-\xi),  \label{Eq_FFT}
\end{equation}
see, e.g., \cite[Section 4.8]{Alberio et al}. We also use the notation $%
\mathcal{F}_{x\rightarrow\kappa}\varphi$ and $\widehat{\varphi}$\ for the
Fourier transform of $\varphi$.

The Fourier transform extends to $L^{2}$. If $f\in L^{2}\left( \mathbb{Q}%
_{p}^{N}\right) $, its Fourier transform is defined as 
\begin{equation*}
(\mathcal{F}f)(\xi)=\lim_{k\rightarrow\infty}\dint \limits_{||x||_{p}\leq
p^{k}}\chi_{p}(\xi\cdot x)f(x)d^{N}x,\quad\text{for }\xi\in%
%TCIMACRO{\U{211a} }%
%BeginExpansion
\mathbb{Q}
%EndExpansion
_{p}^{N},
\end{equation*}
where the limit is taken in $L^{2}\left( \mathbb{Q}_{p}^{N}\right) $. We
recall that the Fourier transform is unitary on $L^{2}\left( \mathbb{Q}%
_{p}^{N}\right) $, i.e. $||f||_{2}=||\mathcal{F}f||_{2}$ for $f\in
L^{2}\left( \mathbb{Q}_{p}^{N}\right) $ and that (\ref{Eq_FFT}) is also
valid in $L^{2}\left( \mathbb{Q}_{p}^{N}\right) $, see, e.g., \cite[Chapter
III, Section 2]{Taibleson}.

\subsection{Distributions}

The $\mathbb{C}$-vector space $\mathcal{D}^{\prime}\left( \mathbb{Q}%
_{p}^{N}\right) $ of all continuous linear functionals on $\mathcal{D}(%
\mathbb{Q}_{p}^{N})$ is called the Bruhat-Schwartz space of distributions.
Every linear functional on $\mathcal{D}(\mathbb{Q}_{p}^{N})$ is continuous,
i.e. $\mathcal{D}^{\prime}\left( \mathbb{Q}_{p}^{N}\right) $\ agrees with
the algebraic dual of $\mathcal{D}(\mathbb{Q}_{p}^{N})$, see, e.g., \cite[%
Chapter 1, VI.3, Lemma]{V-V-Z}.

We endow $\mathcal{D}^{\prime}\left( \mathbb{Q}_{p}^{N}\right) $ with the
weak topology, i.e. a sequence $\left\{ T_{j}\right\} _{j\in\mathbb{N}}$ in $%
\mathcal{D}^{\prime}\left( \mathbb{Q}_{p}^{N}\right) $ converges to $T$ if $%
\lim_{j\rightarrow\infty}T_{j}\left( \varphi\right) =T\left( \varphi\right) $
for any $\varphi\in\mathcal{D}(\mathbb{Q}_{p}^{N})$. The map 
\begin{equation*}
\begin{array}{lll}
\mathcal{D}^{\prime}\left( \mathbb{Q}_{p}^{N}\right) \times\mathcal{D}(%
\mathbb{Q}_{p}^{N}) & \rightarrow & \mathbb{C} \\ 
\left( T,\varphi\right) & \rightarrow & T\left( \varphi\right)%
\end{array}%
\end{equation*}
is a bilinear form which is continuous in $T$ and $\varphi$ separately. We
call this map the pairing between $\mathcal{D}^{\prime}\left( \mathbb{Q}%
_{p}^{N}\right) $ and $\mathcal{D}(\mathbb{Q}_{p}^{N})$. From now on we will
use $\left( T,\varphi\right) $ instead of $T\left( \varphi\right) $.

Every $f$\ in $L_{loc}^{1}$ defines a distribution $f\in\mathcal{D}^{\prime
}\left( \mathbb{Q}_{p}^{N}\right) $ by the formula 
\begin{equation*}
\left( f,\varphi\right) =\dint \limits_{\mathbb{Q}_{p}^{N}}f\left( x\right)
\varphi\left( x\right) d^{N}x.
\end{equation*}

\subsection{\label{SEction Fourier Transform}The Fourier transform of a
distribution}

The Fourier transform $\mathcal{F}\left[ T\right] $ of a distribution $T\in%
\mathcal{D}^{\prime}\left( \mathbb{Q}_{p}^{N}\right) $ is defined by%
\begin{equation*}
\left( \mathcal{F}\left[ T\right] ,\varphi\right) =\left( T,\mathcal{F}\left[
\varphi\right] \right) \text{ for all }\varphi\in\mathcal{D}\left( \mathbb{Q}%
_{p}^{N}\right) \text{.}
\end{equation*}
The Fourier transform $T\rightarrow\mathcal{F}\left[ T\right] $ is a linear
and continuous isomorphism from $\mathcal{D}^{\prime}\left( \mathbb{Q}%
_{p}^{N}\right) $\ onto $\mathcal{D}^{\prime}\left( \mathbb{Q}%
_{p}^{N}\right) $. Furthermore, $T=\mathcal{F}\left[ \mathcal{F}\left[ T%
\right] \left( -\xi\right) \right] $.

Let $T\in\mathcal{D}^{\prime}\left( \mathbb{Q}_{p}^{n}\right) $ be a
distribution. Then supp$T\subset B_{L}^{N}$ if and only if $\mathcal{F}\left[
T\right] $ is a locally constant function, and the exponent of local
constancy of $\mathcal{F}\left[ T\right] $ is $\geq-L$. In addition

\begin{equation*}
\mathcal{F}\left[ T\right] \left( \xi\right) =\left( T\left( y\right)
,\Omega\left( p^{-L}\left\Vert y\right\Vert _{p}\right) \chi_{p}\left(
\xi\cdot y\right) \right) ,
\end{equation*}
see, e.g., \cite[Section 4.9]{Alberio et al}.

\subsection{The direct product of distributions}

Given $F\in\mathcal{D}^{\prime}\left( \mathbb{Q}_{p}^{N}\right) $ and $G\in%
\mathcal{D}^{\prime}\left( \mathbb{Q}_{p}^{M}\right) $, their \textit{direct
product }$F\times G$ is defined by the formula%
\begin{equation*}
\left( F\left( x\right) \times G\left( y\right) ,\varphi\left( x,y\right)
\right) =\left( F\left( x\right) ,\left( G\left( y\right) ,\varphi\left(
x,y\right) \right) \right) \text{ for }\varphi\left( x,y\right) \in\mathcal{D%
}\left( \mathbb{Q}_{p}^{N+M}\right) .
\end{equation*}
The direct product is commutative: $F\times G=G\times F$. In addition the
direct product is continuous with respect to the joint factors.

\subsection{The convolution of distributions}

Given $F,G\in\mathcal{D}^{\prime}\left( \mathbb{Q}_{p}^{N}\right) $, their
convolution $F\ast G$ is defined by%
\begin{equation*}
\left( F\ast G,\varphi\right) =\lim_{k\rightarrow\infty}\left( F\left(
y\right) \times G\left( x\right) ,\Omega\left( p^{-k}\left\Vert y\right\Vert
_{p}\right) \varphi\left( x+y\right) \right)
\end{equation*}
if the limit exists for all $\varphi\in\mathcal{D}\left( \mathbb{Q}%
_{p}^{N}\right) $. We recall that if $F\ast G$ exists, then $G\ast F$ exists
and $F\ast G=G\ast F$, see, e.g., \cite[Section 7.1]{V-V-Z}. If $F,G\in 
\mathcal{D}^{\prime}\left( \mathbb{Q}_{p}^{N}\right) $ and supp$G\subset
B_{L}^{n}$, then the convolution $F\ast G$ exists, and it is given by the
formula%
\begin{equation*}
\left( F\ast G,\varphi\right) =\left( F\left( y\right) \times G\left(
x\right) ,\Omega\left( p^{-L}\left\Vert y\right\Vert _{p}\right)
\varphi\left( x+y\right) \right) \text{ for }\varphi\in\mathcal{D}\left( 
\mathbb{Q}_{p}^{N}\right) .
\end{equation*}
In the case in which $G=\psi\in\mathcal{D}\left( \mathbb{Q}_{p}^{n}\right) $%
, $F\ast\psi$ is a locally constant function given by 
\begin{equation*}
\left( F\ast\psi\right) \left( y\right) =\left( F\left( x\right) ,\psi\left(
y-x\right) \right) ,
\end{equation*}
see, e.g., \cite[Section 7.1]{V-V-Z}.

\subsection{The multiplication of distributions}

Set $\delta_{k}\left( x\right) :=p^{Nk}\Omega\left( p^{k}\left\Vert
x\right\Vert _{p}\right) $ for $k\in\mathbb{N}$. Given $F,G\in\mathcal{D}%
^{\prime}\left( \mathbb{Q}_{p}^{N}\right) $, their product $F\cdot G$ is
defined by%
\begin{equation*}
\left( F\cdot G,\varphi\right) =\lim_{k\rightarrow\infty}\left( G,\left(
F\ast\delta_{k}\right) \varphi\right)
\end{equation*}
if the limit exists for all $\varphi\in\mathcal{D}\left( \mathbb{Q}%
_{p}^{N}\right) $. If the product $F\cdot G$ exists then the product $G\cdot
F$ exists and they are equal.

We recall that \ the existence of the product $F\cdot G$ is equivalent \ to
the existence of $\mathcal{F}\left[ F\right] \ast\mathcal{F}\left[ G\right] $%
. In addition, $\mathcal{F}\left[ F\cdot G\right] =\mathcal{F}\left[ F\right]
\ast\mathcal{F}\left[ G\right] $ and $\mathcal{F}\left[ F\ast G\right] =%
\mathcal{F}\left[ F\right] \cdot\mathcal{F}\left[ G\right] $, see, e.g., 
\cite[Section 7.5]{V-V-Z}.

\begin{acknowledgement}
The author whishes to thank B. A. Zambrano-Luna for his kind assistance in
the elaboration of Figures \ref{Figure 3}-\ref{Figure 4}.
\end{acknowledgement}

\end{document}